\documentclass[aps,pra,floatfix,superscriptaddress,twocolumn,showpacs,showkeys,10pt]{revtex4-1}
\usepackage{amssymb, amsmath,color,mciteplus,graphicx,subfigure}
\usepackage{calc,epsfig,epstopdf,color,mciteplus,bm,mathrsfs}
\usepackage{times}
\usepackage{float}
\usepackage{lipsum}

\begin{document}

\title{Suppression of Universal Errors in DFS-Encoded Superconducting Geometric Logical \emph{T} Gate}

\author{Cheng-Yun Ding}\email{cyding@aqnu.edu.cn}
\affiliation{School of Physics and Astronomy, Anqing Normal University, Anqing 246133, China}

\author{Li-Hua Zhang}\email{zhanglh@aqnu.edu.cn}
\affiliation{School of Electronic Information and Integrated Circuits, Anqing Normal University,
Anqing 246133, China}

\author{Jian Zhou}\email{jianzhou8627@163.com}
\affiliation{School of Electrical and Photoelectronic Engineering, West Anhui University, Lu'an 237012, China}

\date{\today}

\begin{abstract}
High-fidelity logical \emph{T}-gate realization constitutes a core prerequisite for large-scale fault-tolerant quantum computing. However, conventional magic state distillation requires massive physical qubit overhead across successive distillation rounds, alongside sophisticated measurement and feedback control, thereby inducing considerable spatial and temporal resource consumption. Herein, we propose a controlled superconducting geometric logical \emph{T} gate scheme that achieves high-order suppression of universal errors, by integrating decoherence-free subspace encoding with multi-loop optimized composite geometric pulse engineering.  Guided by tailored trajectory design, we systematically establish unified gate construction frameworks for conventional geometric, composite geometric, and optimized composite geometric protocols. By flexibly controling additional parametric degrees of freedom, the proposed scheme achieves substantially enhanced robustness against diverse noise sources. Numerical simulations reveal that, within tunable superconducting quantum circuits, our geometric logical \emph{T} gate outperforms both conventional composite geometric and dynamical gates in suppressing Rabi frequency, detuning, and residual inter-qubit crosstalk errors that can all be suppressed to the fourth order, while additionally providing inherent suppression of collective dephasing errors. The present strategy alleviates intrinsic limitations of mainstream approaches and opens a promising avenue toward robust high-fidelity logical \emph{T} gate construction.

\end{abstract}

\maketitle
\section{Introduction}
Quantum computation \cite{divincenzo1995quantum} is a novel paradigm that leverages quantum coherence and entanglement to achieve exponential parallel acceleration. It demonstrates remarkable efficacy in solving problems--such as quantum random circuit sampling \cite{arute2019quantum}--that are intractable for classical computing. However, quantum state-encoded information is highly susceptible to environmental disturbances, local manipulation errors, and random noise, which degrades qubit coherence and induces irreversible computational errors. Thus, stable, scalable quantum information processing in noisy environments remains the core bottleneck for the practical realization of quantum computation.

Quantum error correction \cite{PhysRevA.55.900} is widely regarded as the fundamental solution to this problem. By encoding multiple physical qubits into one logical qubit and adopting redundant coding for active error detection and correction, such schemes can theoretically suppress error rates polynomially. Nevertheless, due to the high intrinsic noise of physical qubits, conventional quantum error correction protocols require numerous auxiliary physical qubits and complex real-time feedback control. As a result, large resource overhead remains a critical constraint on scalable quantum computing. Notably, fault-tolerant universal quantum computing requires both fault-tolerant Clifford gates and non-Clifford gates, among which the \emph{T} gate is essential. At present, high-fidelity \emph{T}-gate implementation primarily relies on magic state distillation \cite{PhysRevA.71.022316}, which imposes substantial physical qubit overhead during logical distillation procedures \cite{PhysRevA.71.022316,meier2013magic}. Despite extensive advances in optimized error-correction codes \cite{PhysRevA.86.052329,hastings2018distillation,ruiz2026unfolded}, including color codes \cite{lee2025low} and low-density parity-check codes \cite{menon2025magic}, sufficient high-quality physical qubit resources remain inaccessible in intermediate-scale noisy quantum systems. This raises a key question: can we bypass such resource-intensive protocols and develop a streamlined alternative? Instead of realizing all logical gates comprehensively, we may concentrate limited resources to address the bottleneck of \emph{T} gate construction. Direct physical-level generation of high-fidelity logical \emph{T} gate would substantially reshape quantum computing architectures. Furthermore, the threshold theorem \cite{aharonov1997fault} reveals that for a fixed logical error rate, suppressed physical errors can greatly decrease the required number of physical qubits. Therefore, realizing highly fault-tolerant \emph{T} gate with low qubit overhead and simple measurement-feedback operations is vital for scalable fault-tolerant quantum computing.

By virtue of intrinsic fault tolerance, geometric phases \cite{berry1984quantal,PhysRevLett.58.1593} endow the corresponding geometric quantum gates with inherent resilience against local random noises, rendering them a promising paradigm for high-precision quantum manipulation \cite{PhysRevLett.91.090404,PhysRevA.72.020301,PhysRevLett.102.030404}. Early adiabatic geometric quantum computing protocols \cite{ekert2000geometric,zanardi1999holonomic} were severely constrained by finite quantum coherence time and demanding experimental requirements, hindering long-term experimental advancement. Accordingly, nonadiabatic geometric quantum computing (NGQC) schemes using nonadiabatic geometric phases \cite{PhysRevLett.58.1593,anandan1988non} were subsequently developed \cite{PhysRevLett.87.097901,PhysRevLett.89.097902,sjoqvist2012non,PhysRevLett.109.170501}, which enable high-fidelity geometric gate construction via simplified, fast control strategies. Owing to these merits of simplicity, rapidity, and experimental feasibility, NGQC has garnered extensive research interests, with experimental demonstrations accomplished across diverse platforms including superconducting circuits \cite{abdumalikov2013experimental,PhysRevLett.124.230503}, nuclear magnetic resonance systems \cite{feng2013experimental}, and NV centers \cite{zu2014experimental}. Nevertheless, conventional NGQC \cite{PhysRevA.71.014302,PhysRevA.94.052310,PhysRevA.96.052316,PhysRevA.97.022332} imposes stringent constraints on both parallel transport and cyclic evolution, which severely restricts accessible evolution trajectories. This drawback results in longer gate duration compared with dynamical gates and accumulated errors over nonsmooth evolution trajectories, thereby suppressing the full exploitation of intrinsic geometric robustness. To address this limitation, abundant theoretical investigations \cite{PhysRevApplied.14.064009,PhysRevA.109.022613,PhysRevA.80.024302,PhysRevA.103.032609,PhysRevA.109.042615,ding2025composite,PhysRevResearch.2.023295,ding2022path,ding2021nonadiabatic,
li2021high,PhysRevApplied.16.044005,PRXQuantum.2.030333,PhysRevA.106.062402,PhysRevA.108.032616,PhysRevA.109.012619,fn6y-byqc} and experimental validations \cite{PhysRevApplied.12.024024,PhysRevApplied.16.064003,PhysRevLett.127.030502,ai2022experimental,PhysRevApplied.20.054047,PhysRevApplied.19.044076} have been reported, integrating time-optimal control \cite{PhysRevApplied.14.064009,PhysRevA.109.022613}, composite pulses \cite{PhysRevA.80.024302,PhysRevA.103.032609,PhysRevA.109.042615,ding2025composite,PhysRevApplied.12.024024}, shortened evolution paths \cite{PhysRevResearch.2.023295,ding2022path,ding2021nonadiabatic,li2021high,PhysRevApplied.19.044076}, dynamical correction \cite{PhysRevApplied.16.044005,PhysRevApplied.16.064003}, and other optimized control approaches \cite{PRXQuantum.2.030333,PhysRevA.106.062402,PhysRevA.108.032616,PhysRevA.109.012619,fn6y-byqc,PhysRevLett.127.030502,ai2022experimental,PhysRevApplied.20.054047}. Among these strategies, composite pulse engineering stands out as a mainstream technique for enhancing fault tolerance in NGQC \cite{PhysRevLett.124.230503}.
However, conventional composite-pulse geometric gate designs \cite{PhysRevLett.124.230503,PhysRevA.103.032609} only afford robust performance against single-axis parameter errors, such as Rabi frequency or detuning deviations, while remaining highly vulnerable to orthogonal error components. To overcome this deficiency, an optimized composite two-loop NGQC scheme has been theoretically proposed \cite{ding2025composite}. This architecture realizes universal nonadiabatic geometric gates with dual-axis error robustness under a single trajectory configuration, representing a more efficient fault-tolerant manipulation approach. Furthermore, suppressing inter-qubit residual crosstalk errors \cite{PhysRevLett.127.130501,PhysRevLett.127.200502,PhysRevLett.129.060501,PhysRevLett.129.040502} remains another critical issue toward scalable qubit operation and universal quantum computing.

Benefiting from high-precision microwave control compatibility and mature micro-nano fabrication techniques, superconducting quantum circuits \cite{clarke2008superconducting,devoret2013superconducting,kjaergaard2020superconducting,huang2020superconducting} feature flexible, precise manipulation and excellent scalability, rendering them one of the most promising physical platforms for quantum computing. For capacitively coupled transmon qubits \cite{devoret2013superconducting,kjaergaard2020superconducting,huang2020superconducting,PhysRevA.76.042319}, however, the intrinsic coupling strength is generally fixed once device parameters are determined, which impedes the development of stable and controllable superconducting quantum computing. To address this issue, tunable inter-qubit coupling can be achieved via an ac magnetic flux driving \cite{reagor2018demonstration,PhysRevApplied.10.054009,PhysRevLett.123.080501,PhysRevApplied.13.064012}, which enables controllable quantum manipulation. In addition, decoherence-free subspace (DFS) encoding \cite{PhysRevLett.79.1953,PhysRevLett.79.3306,PhysRevLett.81.2594,kwiat2000experimental} represents an effective fault-tolerant coding strategy, where at least two physical qubits are utilized to construct a single logical qubit. By exploiting symmetric system-environment interactions, DFS embeds quantum information into noise-immune subspaces, intrinsically suppressing collective dephasing errors. Nevertheless, on practical superconducting chips, intrinsic parameter mismatches in frequency and anharmonicity inevitably exist between adjacent qubits. Such minor asymmetry readily destroys the ideal subspace structure and induces undesired quantum state leakage. Leveraging the tunable coupling of transmon systems, external microwave driving parameters can be precisely calibrated to suppress leakage errors effectively, thereby enabling near-ideal realization of DFS encoding.

In this paper, we implement the controlled fourth-order suppression of universal errors in nonadiabatic non-Clifford geometric logical \emph{T} gate based on a superconducting circuit with tunable coupling, leveraging parsimonious DFS encoding and optimized multi-loop geometric composite pulse technology. Numerical results demonstrate that the proposed superconducting geometric logical \emph{T} gate not only suppresses collective dephasing errors but also efficiently mitigates omnidirectional error sources, including Rabi, detuning, and crosstalk errors. Compared with magic state distillation-based active error correction schemes, our approach offers two key advantages: (1) Employing DFS encoding within superconducting tunable architecture, our scheme inherently immunizes against collective dephasing errors and features low resource overhead. It eliminates the need for complex measurement-feedback operations to stabilize quantum information, significantly simplifying experimental implementation and enhancing system stability. Meanwhile, the leakage error from high-frequency oscillations can also be effectively suppressed. (2) The optimized geometric multi-loop composite pulse technology maximally suppresses the universal errors, including Rabi, detuning, and crosstalk errors, via flexible and controllable trajectory correction. The hybrid protection strategy integrating these two fault-tolerant methods may provides a novel pathway for realizing hardware-efficient, stong-robust logical \emph{T} gate against universal errors.

\section{Physical model with DFS encoding}\label{sec1}
In this section, we first derive the effective Hamiltonian for two capacitively coupled transmon qubits arranged in a two-dimensional (2D) superconducting square lattice. This configuration supports a single-excitation 2D DFS, which inherently immunizes the encoded quantum information against collective dephasing noise. We then characterize the primary noise sources encountered in the manipulation of superconducting qubits for gate implementation within our theoretical framework.
\subsection{Effective Hamiltonian between Two Transmons}\label{sec0}
\begin{figure*}[tbp]
\centering
\includegraphics[width=0.95\linewidth]{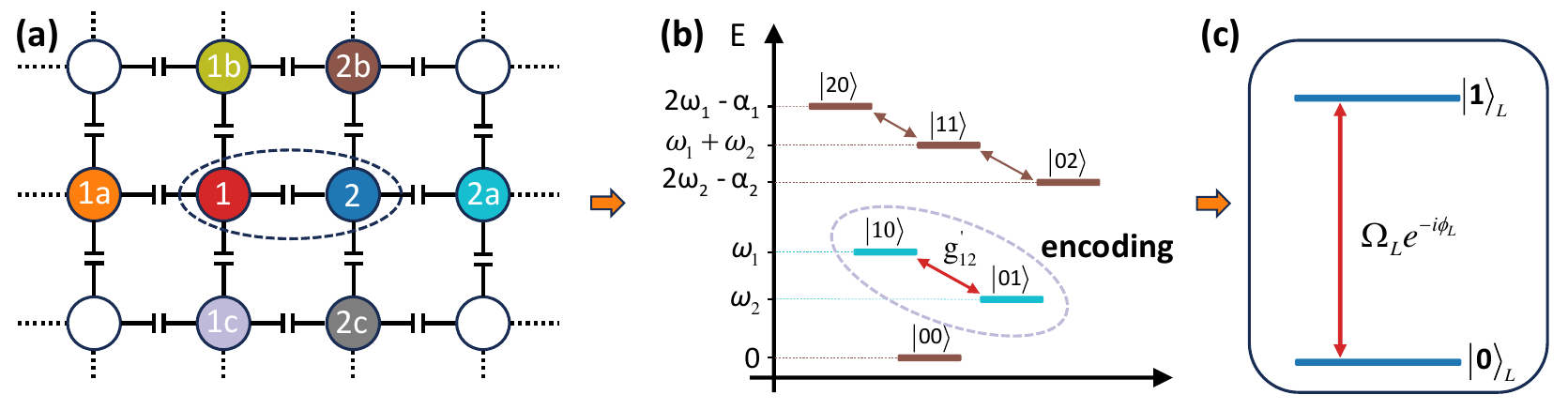}
\caption{Illustration of our superconducting implementation with DFS encoding. (a) Schematic of the 2D square transmon lattice via capacitive coupling. Transmons $1$ and $2$ are designated as target qubits, and surrounding devices ($1a$, $1b$, $1c$, $2a$, $2b$, $2c$) act as spectator qubits, with different colors representing disparate resonant frequencies. (b) Energy-level spectrum of target qubits $1$ and $2$. The single-excitation transition subspace naturally forms a feasible DFS for single-logical-qubit encoding. (c) Encoded two-level logical basis structure corresponding to $|10\rangle=|0\rangle_\mathrm{L}$ and $|01\rangle=|1\rangle_\mathrm{L}$, where resonant transitions can be flexibly tuned via the effective coupling strength $\Omega_\mathrm{L}$.
}\label{Fig1}
\end{figure*}

As shown in Fig. \ref{Fig1}(a), we select two adjacent transmons, labeled $1$ and $2$, as our target system. With fixed resonant frequencies and purely capacitive coupling, the coupling strength is generally non-tunable, which constrains the realization of high-performance quantum gates. To address this limitation, we adopt a parametrically tunable coupling model, widely validated experimentally \cite{reagor2018demonstration,PhysRevApplied.10.054009,PhysRevLett.123.080501,PhysRevApplied.13.064012}, by applying an additional ac microwave driving to one of the transmons. Specifically, considering up to the two-excitation subspace and setting $\hbar=1$ hereafter, the original Hamiltonian for these two transmons can be expressed as $H_1=H_0+V$, where the free Hamiltonian $H_0$ and interaction Hamiltonian $V$ are given by, respectively,
\begin{eqnarray}
H_0&=&\sum_{j=1}^{2}\sum_{k=1}^{2}\left[k\omega_j-\frac{k(k-1)}{2}\alpha_j\right]|k\rangle_j\langle k|, \nonumber \\
V&=&g_{12}\prod_{j=1,2}\left(\sum_{k=1}^{2}\lambda_j|k-1\rangle_j\langle k|\right),
\end{eqnarray}
in which $\lambda_1=1$ and $\lambda_2=\sqrt{2}$; $g_{12}$ denotes the coupling strength between the two transmons; $\omega_j$ and $\alpha_j$ represent the transition frequency and anharmonicity of the $j$-th transmon, respectively. Performing transformation to the interaction picture via the rotation operator $U_1=\exp(-\mathrm{i}H_0t)$, the original Hamiltonian $H_1$ is transformed into
\begin{align}
H_1^I&=\mathrm{i}\frac{\mathrm{d}U_1^{\dagger}}{\mathrm{d}t}+U_1^{\dagger}H_1U_1 \nonumber \\
&=g_{12}\Big\{|10\rangle_{12}\langle01|e^{\mathrm{i}\Delta t}+\sqrt{2}|20\rangle_{12}\langle11| \nonumber  \\
&\quad\times e^{\mathrm{i}(\Delta-\alpha_1)t}+\sqrt{2}|11\rangle_{12}\langle02|e^{\mathrm{i}(\Delta+\alpha_2)t}+\textrm{H.c.} \Big\}.
\end{align}
Here, we define $\Delta = \omega_1-\omega_2$, and neglect fast-oscillating terms that violate energy conservation. Furthermore, once transmon parameters are fixed, their equivalent coupling strength within each transition subspace remains constant, limiting further optimization. To enable time-dependent modulation of the coupling strength, we apply an additional ac microwave driving to transmon $2$, rendering its transition frequency time-dependent: $\omega_2(t)=\omega_2+\varepsilon(t)$, where
\begin{eqnarray}
\varepsilon(t)=\dot{f}(t), \quad f(t)=-\beta\cos{(\nu t+\phi)}.
\end{eqnarray}
Thus, following frequency modulation, the total system Hamiltonian can be rewritten as
\begin{eqnarray}
H_2&=&H_1+H_0'  \nonumber \\
&=&H_0+V+\varepsilon(t)\sum_{k=1}^2k|k\rangle_2\langle k|.
\end{eqnarray}
To clarify the dynamics of the interaction Hamiltonian, we perform a new picture transformation using the operator $U_3=U_2U_1$, given by
\begin{eqnarray}
U_1&=&e^{-\mathrm{i}H_0t}, \nonumber \\
U_2&=&e^{-\mathrm{i}\int H_0'\mathrm{d}t}=e^{-\mathrm{i}f(t)\sum_{k=1}^2k|k\rangle_2\langle k|}.
\end{eqnarray}
Using several commutation relations, the transformed Hamiltonian is derived as
\begin{eqnarray}\label{eq6}
H_2^I&=&\mathrm{i}\frac{\mathrm{d}U_3^{\dagger}}{\mathrm{d}t}+U_3^{\dagger}H_2U_3=H_1^Ie^{-\mathrm{i}f(t)} \nonumber \\
&=&H_1^Ie^{\mathrm{i}\beta\cos(\nu t+\phi)}.
\end{eqnarray}

Next, we apply a reasonable approximation to derive the target DFS. Utilizing the Jacobi-Anger identity
\begin{eqnarray}
e^{\mathrm{i}\beta\cos(\nu t+\phi)}=\sum\nolimits_{m=-\infty}^{+\infty}\mathrm{i}^mJ_m(\beta)e^{\mathrm{i} m(\nu t+\phi)},
\end{eqnarray}
we rewrite Eq. (\ref{eq6}) as
\begin{align}
H_2^I&=\sum_{m=-\infty}^{+\infty}\mathrm{i}^mJ_m(\beta)g_{12}\Big\{\big[|10\rangle_{12}\langle01|e^{\mathrm{i}\Delta t}+\sqrt{2}|20\rangle_{12}\langle11| \nonumber \\
&\quad\times e^{\mathrm{i}(\Delta-\alpha_1)t}+\sqrt{2}|11\rangle_{12}\langle02|e^{\mathrm{i}(\Delta+\alpha_2)t}\big]e^{\mathrm{i}m(\nu t+\phi)}\Big\} \nonumber \\
&\quad+\textrm{H.c.},
\end{align}
where $J_m(\beta)$ is the Bessel function of the first kind and \textrm{H.c.} denotes the Hermitian conjugate. When the ac microwave driving frequency satisfies the following distinct resonance conditions:
\begin{equation}
\left\{
     \begin{array}{lr}
      \nu=\Delta/n_1, \quad \qquad\quad\,  n_1=\pm1,\pm2,..., \\
      \nu=(\Delta-\alpha_1)/n_2,\quad n_2=\pm1,\pm2,..., \\
      \nu=(\Delta+\alpha_2)/n_3,\quad n_3=\pm1,\pm2,...,
    \end{array}
\right.
\end{equation}
different energy-level transitions can be induced within the respective subspaces $\{|10\rangle_{12},|01\rangle_{12}\}$, $\{|11\rangle_{12},|20\rangle_{12}\}$, and $\{|11\rangle_{12},|02\rangle_{12}\}$; the corresponding energy spectrum is shown in Fig. \ref{Fig1}(b). Moreover, setting the driving frequency equal to the transition frequency difference between the two transmon qubits ($\nu=\Delta$) results in resonant interaction for $m=-1$, with all other terms behaving as high-frequency oscillations within the single-excitation subspace $\{|10\rangle_{12},|01\rangle_{12}\}$. Neglecting these high-frequency oscillating terms, the effective Hamiltonian within this subspace is given by
\begin{eqnarray}\label{eq9}
H_{\mathrm{E}}=g'_{12}\left[|10\rangle_{12}\langle01|e^{-\mathrm{i}(\phi-\frac{\pi}{2})}+\textrm{H.c.}\right],
\end{eqnarray}
with effective coupling strength $g'_{12}=J_1(\beta)g_{12}$, whose magnitude can be flexibly tuned by parameter $\beta$. When quantum information is encoded within the two lowest energy levels of transmons, denoted $|0\rangle$, $|1\rangle$, the single-excitation subspace $\{|10\rangle_{12},|01\rangle_{12}\}$ constitutes a feasible DFS. Accordingly, we encode one logical qubit in this subspace via the mapping $|10\rangle_{12}=|0\rangle_\mathrm{L}$ and $|01\rangle_{12}=|1\rangle_\mathrm{L}$. Under this logical basis, Eq. (\ref{eq9}) is transformed into
\begin{eqnarray}\label{eq10}
H_\mathrm{L}=\frac{\Omega_\mathrm{L}}{2}\left(
      \begin{array}{cc}
        0& e^{-\mathrm{i}\phi_\mathrm{L}} \\
        e^{-\mathrm{i}\phi_\mathrm{L}} & 0 \\
      \end{array}
    \right),
\end{eqnarray}
where $\Omega_\mathrm{L}=2g'_{12}$ and $\phi_\mathrm{L}=\phi-\pi/2$. This Hamiltonian corresponds to a general resonant $2$D Hamiltonian, which enables the implementation of arbitrary single-logical-qubit gates. The corresponding encoded energy-level structure is illustrated in Fig. \ref{Fig1}(c).

\subsection{Noise Sources}\label{sec2}
In the 2D superconducting transmon circuit shown in Fig. \ref{Fig1}(a), transmons $1$ and $2$ are selected as target qubits, while their six surrounding devices ($1a$, $1b$, $1c$, $2a$, $2b$, $2c$) serve as spectator qubits. During the implementation of logical-qubit gates via ac microwave-driven tunable inter-qubit coupling on qubit $2$, multiple intrinsic noise channels emerge naturally. Since we adopt static coupling strengths instead of time-dependent microwave pulses for quantum control, errors originating from imperfect pulse shapes and finite bandwidth limitations are safely negligible. Nevertheless, two dominant error mechanisms persist, namely calibration errors and residual ZZ crosstalk error.

Calibration errors originate from two primary contributions. The first is Rabi frequency error, arising from inaccuracies in microwave amplitude calibration. This error can be phenomenologically modeled as
\begin{eqnarray}
g_{12}\longrightarrow g_{12}+\epsilon',
\end{eqnarray}
where $\epsilon'$ denotes the corresponding error amplitude. In conjunction with the geometric gate construction presented in the subsequent section, such deviations disrupt the duration of each gate segment, preventing the parametric evolution trajectory from returning to its initial state to complete cyclic evolution, and thereby degrading overall gate fidelity. The second contribution is detuning error, induced by random frequency drifts of qubit $1$ during flux driving on qubit $2$. This error is correspondingly modeled as
\begin{eqnarray}
\Delta\longrightarrow \Delta+\delta',
\end{eqnarray}
where $\delta'$ represents the associated detuning error amplitude.

Furthermore, ZZ crosstalk constitutes an intrinsic limitation for large-scale quantum computing based on integrated superconducting qubit chips. In our 2D lattice architecture, gate operations and tunable interactions between target qubits $1$ and $2$ induce unintended coupling to surrounding spectator qubits, resulting in undesirable quantum state leakage. This effect originates from the non-negligible, uncontrolled inter-qubit coupling between target and spectator devices. Analogous parasitic coupling persists when implementing additional gate sets on other qubit pairs, giving rise to accumulated and cascaded error propagation across superconducting circuit. Although such crosstalk can be experimentally mitigated via large frequency detuning between target and spectator qubits, weak residual interactions still persist, namely residual ZZ crosstalk \cite{PhysRevLett.127.130501,PhysRevLett.127.200502,PhysRevLett.129.060501,PhysRevLett.129.040502}, which remains a prominent barrier toward scalable fault-tolerant superconducting quantum computing. For the target qubit pair $1$ and $2$, the noise Hamiltonian induced by residual ZZ crosstalk is formulated as
\begin{eqnarray}
H_{\mathrm{zz}}=\sum_{i=1,2}\sum_{j=a,b,c}\frac{\eta_{ij}}{2}\sigma_\mathrm{z}^{(i)}\otimes\sigma_\mathrm{z}^{(ij)},
\end{eqnarray}
where $\eta_{ij}$ represents the coupling strength between target qubit $i$ and spectator qubit $ij$, and $\sigma_\mathrm{z}^{(i)}$ and $\sigma_\mathrm{z}^{(ij)}$ denote the corresponding Pauli-$Z$ operators. In the subsequent section, we elaborate on how the integration of nonadiabatic geometric phases and composite pulse engineering enables the construction of strongly robust nonadiabatic geometric non-Clifford logical \emph{T} gate, which simultaneously suppress all three aforementioned noise channels.

\section{Super-robust nonadiabatic geometric \emph{T} gate}
In this section, we first establish a universal path-design framework for constructing arbitrary single-qubit quantum gates, and subsequently derive the target non-Clifford \emph{T} gate under prescribed boundary conditions. This framework further enables the construction of geometric \emph{T} gates corresponding to three distinct evolution trajectory strategies: conventional orange-sliced shaped, composite, and optimized composite protocols. Second, we numerically characterize the gate robustness with respect to the three dominant noise channels elaborated in Sec. \ref{sec2}. Numerical simulations reveal that the conventional composite geometric scheme only delivers limited resilience against these noise sources. By contrast, our optimized composite scheme achieves markedly enhanced robustness toward each individual noise, and simultaneously suppresses multiple coexisting noise channels along a single optimal evolution path. Finally, we further compare the gate performance of representative schemes under environmental decoherence, which corroborates the superior performance of our optimized composite strategy.

\subsection{General Framework and Gate Construction}\label{sec3}
For a driven qubit system, the general Hamiltonian is given by
\begin{eqnarray}\label{eq15}
H(t)=\frac{1}{2}\left(
               \begin{array}{cc}
                 -\Delta(t) & \Omega(t)e^{-\mathrm{i}\phi(t)} \\
                 \Omega(t)e^{\mathrm{i}\phi(t)} & \Delta(t) \\
               \end{array}
             \right),
\end{eqnarray}
in the computational basis $\{|0\rangle, |1\rangle\}$, where $\Delta(t)$, $\Omega(t)$ and $\phi(t)$ denote the time-dependent detuning, amplitude and phase of the driving field, respectively. To construct the target geometric gates, we introduce a set of evolution states $|\varphi_l(t)\rangle=e^{\mathrm{i}\gamma_l(t)}|a_l(t)\rangle$ (for $l=1,2$) that satisfy the Schr\"{o}dinger equation. Here, $\gamma_l(0)=0$, $\gamma_l(t)$ represents the accumulated global phase. The set of orthogonal auxiliary bases $\{|a_l(t)\rangle\}$ is expressed as
\begin{equation}
\left\{
     \begin{array}{lr}
     |a_1(t)\rangle=\cos\frac{\alpha(t)}{2}|0\rangle+\sin\frac{\alpha(t)}{2}e^{\mathrm{i}\beta(t)}|1\rangle, &  \\
                                          \\
     |a_2(t)\rangle=\sin\frac{\alpha(t)}{2}e^{-\mathrm{i}\beta(t)}|0\rangle-\cos\frac{\alpha(t)}{2}|1\rangle, & \\
     \end{array}
\right.
\end{equation}
whose dynamics can be parameterized by time-dependent polar angle $\alpha(t)$ and azimuthal angle $\beta(t)$ on the Bloch sphere, without directly satisfying the Schr\"{o}dinger equation. By solving the Schr\"{o}dinger equation $\mathrm{i}|\dot{\varphi}_l(t)\rangle=H(t)|\varphi(t)\rangle$ associated with the evolution states $|\varphi_l(t)\rangle$, we establish a direct correspondence between the Hamiltonian parameters $\{\Delta(t),\Omega(t),\phi(t)\}$ and the evolution trajectory parameters $\{\alpha(t),\beta(t)\}$, namely,
\begin{equation}\label{eq17}
\left\{
     \begin{array}{lr}
     \dot{\alpha}(t)=\Omega(t)\sin[\phi(t)-\beta(t)],  &  \\
     \dot{\beta}(t)=-\Delta(t)-\Omega(t)\cot\alpha(t)\cos[\phi(t)-\beta(t)]. & \\
     \end{array}
\right.
\end{equation}
This relation demonstrates that $\alpha(t)$ and $\beta(t)$ of arbitrary form can be achieved by specifying the functional forms of the Hamiltonian parameters $\{\Delta(t),\Omega(t),\phi(t)\}$. Conversely, by designing a desired evolution trajectory, the corresponding Hamiltonian parameters can be inversely derived using this correlation. We impose a cyclic evolution condition with period $\tau$ on the auxiliary states, requiring $|a_l(\tau)\rangle=|a_l(0)\rangle=|\varphi_l(0\rangle)$. This condition generates boundary constraints on the evolution trajectory, given by
\begin{eqnarray}
\alpha(\tau)=\alpha(0)=\alpha_0; \quad\beta(\tau)=\beta(0)\pm 2k\pi=\beta_0,
\end{eqnarray}
with $k=0,1,2,...$. Accordingly, the corresponding evolution operator is derived as
\begin{eqnarray}
U(\tau)&=&\sum_{l=1}^{2}|\varphi_l(\tau)\rangle\langle\varphi_l(0)| \nonumber\\
&=&\sum_{l=1}^{2}e^{\mathrm{i}\gamma_l(\tau)}|a_l(0)\rangle\langle a_l(0)|=e^{-\mathrm{i}\gamma\boldsymbol{n}\cdot\boldsymbol{\sigma}}.
\end{eqnarray}
Here, $\gamma=-\gamma_1(\tau)=\gamma_2(\tau)$ represents the total accumulated phase over the entire evolution. This total phase can be decomposed into $\gamma=\gamma_d+\gamma_g$, where $\gamma_d$ is the dynamical phase and $\gamma_g$ denotes the geometric phase. Their respective integral expressions are given by
\begin{eqnarray}\label{eq20}
\gamma_d&=&\int_{0}^{\tau}\langle \varphi_1(t)|H(t)|\varphi_1(t)\rangle \mathrm{d}t \nonumber\\
&=&-\int_{0}^{\tau}\frac{\dot{\beta}(t)\sin^2\alpha(t)+\Delta(t)}{2\cos\alpha(t)}\mathrm{d}t,
\end{eqnarray}
and
\begin{eqnarray} \label{eq21}
\gamma_g&=&-\mathrm{i}\int_{0}^{\tau}\langle \varphi_1(t)|\partial/\partial t|\varphi_1(t)\rangle \mathrm{d}t \nonumber\\
&=&\frac{1}{2}\int_{0}^{\tau}\dot{\beta}(t)[1-\cos\alpha(t)]\mathrm{d}t.
\end{eqnarray}
Note that $\boldsymbol{n}=(\sin\alpha_0\cos\beta_0$, $\sin\alpha_0\sin\beta_0,\cos\alpha_0)$ and $\boldsymbol{\sigma}=(\sigma_x,\sigma_y,\sigma_z)$ denote an arbitrary unit vector and the well-known Pauli vector, respectively. Thus, $U(\tau)$ corresponds to a rotation gate about the $\boldsymbol{n}$-axis with a rotation angle of $2\gamma$. By specifying the total phase $\gamma$ and the boundary conditions $(\alpha_0,\beta_0)$ of the evolution states  as
\begin{eqnarray}
\gamma=\pi/8,\quad \alpha_0=\beta_0=0,
\end{eqnarray}
we obtain a phase gate that rotates by $\pi/4$ about the $\mathrm{z}$-axis, known as the non-Clifford \emph{T} gate.  This configuration corresponds to the starting point of the evolution state $\varphi_1(t)$ being located at the North Pole of the Bloch sphere. Furthermore, since the functional forms of $\alpha(t)$ and $\beta(t)$ are unconstrained, various closed evolution trajectories of distinct shapes can be chosen to implement the same target \emph{T} gate.

\begin{figure}[tbp]
\centering
\includegraphics[width=0.98\linewidth]{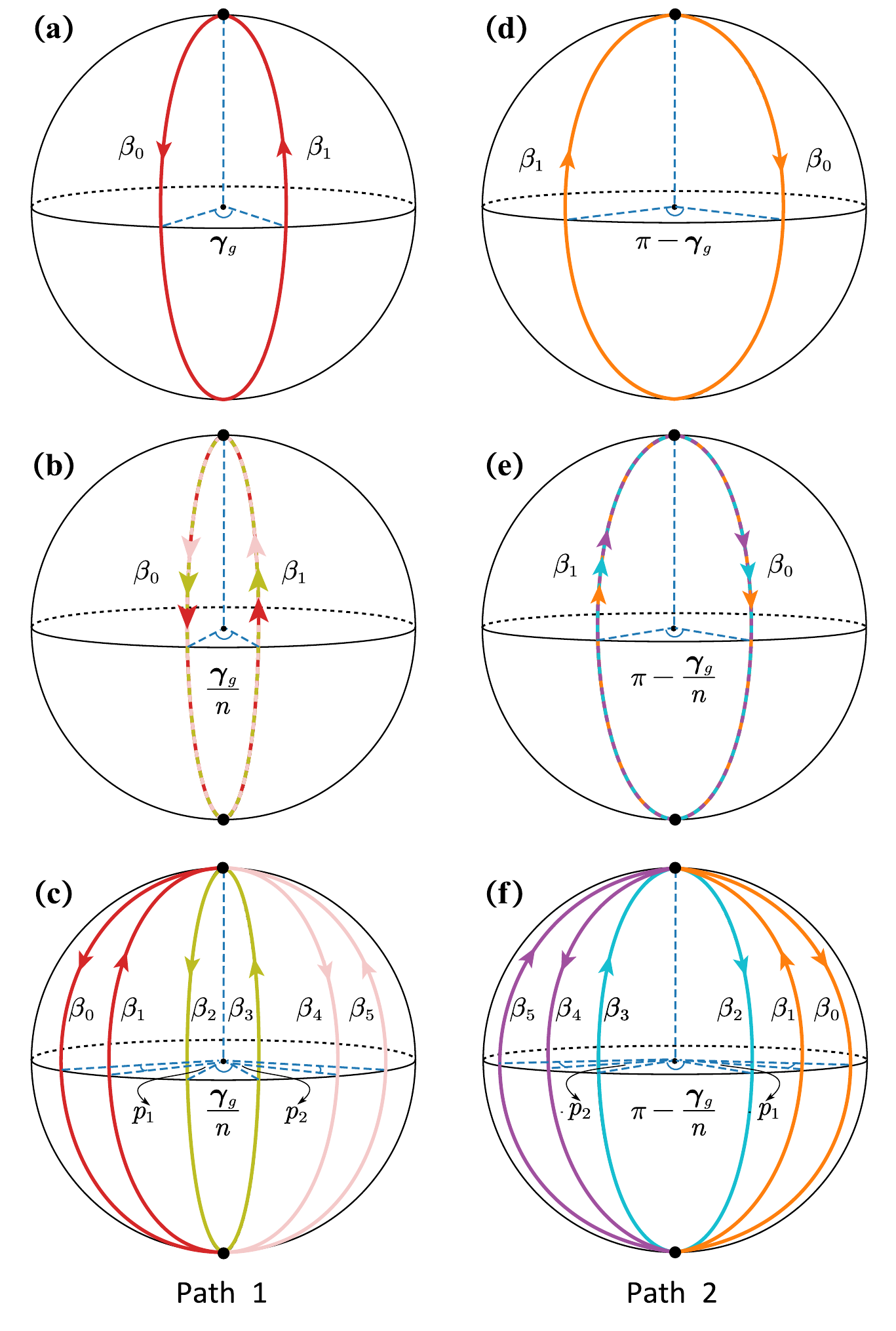}
\caption{Evolution trajectories for different geometric \emph{T} gate schemes. (a)-(c) depict the trajectory strategies of the conventional two-segment orange-sliced shaped single-loop, standard composite $n$-loop, and our optimized composite $n$-loop schemes, corresponding to Path $1$. (d)-(f) show the corresponding strategies for Path $2$. For clarity, only up to three loops are illustrated, where $p_1$ and $p_2$ represent the tunable degrees of freedom for parameter optimization.
}\label{Fig2}
\end{figure}

As evident from Eqs. (\ref{eq20}) and (\ref{eq21}), the geometric phase equals half the solid angle subtended by the corresponding evolution trajectory, which encodes the global geometric property of the evolution. By contrast, the dynamical phase relies on the detailed dynamical evolution along the trajectory. Accordingly, purely geometric quantum gates constructed exclusively from geometric phases exhibit intrinsic noise resilience. Notably, for identical target geometric gates, their robustness performance generally varies across distinct evolution trajectories \cite{ding2022path}. To eliminate the accompanying dynamical phase and satisfy the parallel-transport condition $\langle\varphi_l(t)|H(t)|\varphi_l(t)\rangle=0$, a conventional strategy adopts the parameter setting $\Delta(t)\equiv0$ and $\phi(t)=\beta(t)\pm\pi/2$ for constructing target geometric \emph{T} gate. Substituting into Eq. (\ref{eq17}), this condition imposes $\alpha(\tau)=\alpha(0)\pm\int_{0}^{\tau}\Omega(t)\mathrm{d}t$ and $\dot{\beta}(t)=0$. This constraint indicates that all evolution trajectories are confined purely to longitudinal lines on the Bloch sphere. Such trajectories enable straightforward experimental implementation with time-independent driving detuning and phase. In the following, we analyze three distinct longitudinal-line path strategies, and reconstruct the corresponding gate construction framework from a trajectory-design perspective.

First, we consider the two-segment single-loop orange-sliced shaped geometric \emph{T} gate. As illustrated in Fig. \ref{Fig2}(a), its evolution trajectory comprises two distinct longitudinal lines. Furthermore, two independent closed paths can be adopted to implement this identical geometric \emph{T} gate, denoted as \textbf{Path} $\mathbf{1}$ and \textbf{Path} $\mathbf{2}$. The polar and azimuthal angles for the cyclic Path $1$ are explicitly designed as
 \begin{equation}\label{eq23}
\left\{
     \begin{array}{lr}
\alpha(0)=0\;\;\rightsquigarrow\quad\;\;\;\,\!\alpha(\tau_1)=\pi\;\;\:\rightsquigarrow\alpha(\tau)=0, \\
\beta(0)=\beta_0\rightarrow \begin{array}{c}
                              \beta(\tau_1-\varepsilon)=\beta_0 \\
                              \beta(\tau_1+\varepsilon)=\beta_1
                            \end{array}\rightarrow\beta(\tau)=\beta_1,
     \end{array}
\right.
\end{equation}
where the arrows ``$\rightsquigarrow$" and ``$\rightarrow$" denote whether the parameters $\alpha(t)$ and $\beta(t)$ vary or stay constant over time during the evolution process, respectively. $\varepsilon\ll0$ represents the instantaneous jump time, at which the phase parameter switches from $\beta_0$ to $\beta_1$ at intermediate time $\tau_1$. In other words, the overall evolution proceeds from the North Pole to the South Pole along the longitudinal line $\beta_0$, and subsequently returns to the initial starting point along a separate longitudinal line $\beta_1$. Notably, since the initial state resides at the North Pole, distinct values are permissible for the boundary azimuthal angles $\beta(0)$ and $\beta(\tau)$. In addition, a parameter discontinuity occurs at the South Pole. Combining with Eq. (\ref{eq17}) and Eq. (\ref{eq23}), the corresponding piecewise Hamiltonian parameters are inversely derived as
\begin{equation}\label{eq24}
\left\{
     \begin{array}{lr}
     \int_{0}^{\tau_1}\Omega(t)\mathrm{d}t=\pi, \quad\phi(t)=\beta_0+\frac{\pi}{2}, \quad t\in[0,\tau_1],  &  \\
     \\
     \int_{\tau_1}^{\tau}\Omega(t)\mathrm{d}t=\pi,\quad\phi(t)=\beta_1-\frac{\pi}{2},\quad t\in[\tau_1,\tau]. & \\
     \end{array}
\right.
\end{equation}
The angular separation between these two longitudinal lines directly corresponds to the geometric phase, namely $\gamma_g=\beta_1-\beta_0$. This relation is verified by computing half the solid angle subtended by Path $1$. The resulting overall evolution operator is given by
\begin{eqnarray}
U_1(\tau)=U(\tau,\tau_1)U(\tau_1,0)=e^{-\mathrm{i}\gamma_g\sigma_\mathrm{z}},
\end{eqnarray}
which corresponds to the target geometric \emph{T} gate with $\gamma_g=\pi/8$. For Path $2$, the trajectory and Hamiltonian parameters are constructed to follow the forms presented in Eqs. (\ref{eq23}) and (\ref{eq24}), with the only modification being $\beta_1-\beta_0=\gamma_g-\pi$, as depicted in Fig. \ref{Fig2}(d). Accordingly, the corresponding evolution operator reads $U_2(\tau)=e^{-\mathrm{i}(\gamma_g-\pi)\sigma_\mathrm{z}}=-e^{-\mathrm{i}\gamma_g\sigma_\mathrm{z}}$, which is equivalent to the evolution operator $U_1(\tau)$ derived for Path $1$.

Second, we introduce the conventional multi-loop composite geometric \emph{T} gate. Its evolution trajectory is constructed by repeating the single-loop structure of the first strategy for $n$ cycles, as depicted in Figs. \ref{Fig2}(b) and \ref{Fig2}(e). Since the fundamental single-loop scheme possesses two independent paths, this composite one likewise supports two distinct evolution paths. The parameter designs for Path $1$ and Path $2$ are given by
\begin{widetext}
\begin{eqnarray}
\left\{
\begin{array}{lr}
\alpha(\tau_0)=0\,\:\rightsquigarrow\quad\quad\alpha(\tau_1)=\pi\,\:\;\rightsquigarrow\quad\quad\alpha(\tau_2)=0\quad\!\rightsquigarrow\;...\;\rightsquigarrow \quad\quad\alpha(\tau_{2n-1})=\pi\;\;\rightsquigarrow\alpha(\tau_{2n})=0,   \\
\beta(\tau_0)=\beta_0\rightarrow\begin{array}{c}
                              \beta(\tau_1-\varepsilon)=\beta_0 \\
                              \beta(\tau_1+\varepsilon)=\beta_1
                            \end{array}\rightarrow
                            \begin{array}{c}
                              \beta(\tau_2-\varepsilon)=\beta_1 \\
                              \beta(\tau_2+\varepsilon)=\beta_0
                            \end{array}  \rightarrow \;...\;
\rightarrow\begin{array}{c}
                              \beta(\tau_{2n-1}-\varepsilon)=\beta_0 \\
                              \beta(\tau_{2n-1}+\varepsilon)=\beta_1
                              \end{array} \rightarrow\beta(\tau_{2n})=\beta_1,
\end{array}
\right.
\end{eqnarray}
\end{widetext}
where $\tau_0=0$, $\tau_{2n}=\tau$. The angular difference satisfies $\beta_1-\beta_0=\gamma_g/n$ for Path $1$, and $\beta_1-\beta_0=\gamma_g/n-\pi$ for Path $2$, respectively. To synthesize the target geometric \emph{T} gate, the Hamiltonian parameters are partitioned into $n$ identical segments following the parameter form presented in Eq. (\ref{eq24}). Accordingly, the overall composite evolution operator is derived as

\begin{eqnarray}\label{eq27}
U_c^m(\tau)&=&U(\tau,\tau_{2n-1})...U(\tau_2,\tau_1)U(\tau_1,0) \nonumber \\
&=&\left[(-1)^{m+1}e^{-\mathrm{i}\gamma_g\sigma_\mathrm{z}/n}\right]^n \nonumber \\
&=&(-1)^{n(m+1)}e^{-\mathrm{i}\gamma_g\sigma_\mathrm{z}}=T.
\end{eqnarray}

Here, \emph{T} denotes the target geometric \emph{T} gate, and $m=1,2$ label Path $1$ and Path $2$, respectively, with the preset geometric phase $\gamma_g=\pi/8$.

Third, we present our optimized multi-loop composite geometric \emph{T} gate. Distinct from the conventional composite \emph{T}-gate strategy in which every orange-sliced shaped loop possesses identical inter-loop angular separations, our proposed scheme permits asymmetric angular intervals between adjacent loops. Specifically, the parameter designs for Path $1$ and Path $2$ are given by

\begin{widetext}
\begin{eqnarray}
\left\{
\begin{array}{lr}
\alpha(\tau_0)=0\,\:\rightsquigarrow\quad\quad\alpha(\tau_1)=\pi\,\:\;\rightsquigarrow\quad\quad\alpha(\tau_2)=0\quad\!\rightsquigarrow\;...\;\rightsquigarrow \quad\quad\alpha(\tau_{2n-1})=\pi\;\;\quad\;\;\rightsquigarrow\alpha(\tau_{2n})=0,    \\
\beta(\tau_0)=\beta_0\rightarrow\begin{array}{c}
                              \beta(\tau_1-\varepsilon)=\beta_0 \\
                              \beta(\tau_1+\varepsilon)=\beta_1
                            \end{array}\rightarrow
                            \begin{array}{c}
                              \beta(\tau_2-\varepsilon)=\beta_1 \\
                              \beta(\tau_2+\varepsilon)=\beta_2
                            \end{array}  \rightarrow \;...\;
\rightarrow\begin{array}{c}
                              \beta(\tau_{2n-1}-\varepsilon)=\beta_{2n-2} \\
                              \beta(\tau_{2n-1}+\varepsilon)=\beta_{2n-1}
                              \end{array} \rightarrow\beta(\tau_{2n})=\beta_{2n-1},
\end{array}
\right.
\end{eqnarray}
\end{widetext}
and the corresponding Hamiltonian parameters are inversely determined as
\begin{eqnarray}
\left\{
  \begin{array}{ll}
    \int_{0}^{\tau_1}\Omega(t)\mathrm{d}t=\pi,  &  \hbox{$\phi(t)=\beta_0+\frac{\pi}{2}$, $t\in[0,\tau_1]$;} \\
            \\
    \int_{\tau_1}^{\tau_2}\Omega(t)\mathrm{d}t=\pi, & \hbox{$\phi(t)=\beta_1-\frac{\pi}{2}$, $t\in[\tau_1,\tau_2]$;} \\
           \\
    \int_{\tau_2}^{\tau_3}\Omega(t)\mathrm{d}t=\pi, & \hbox{$\phi(t)=\beta_2+\frac{\pi}{2},t\in[\tau_2,\tau_3]$;} \\
           \\
    \int_{\tau_3}^{\tau_4}\Omega(t)\mathrm{d}t=\pi, & \hbox{$\phi(t)=\beta_3-\frac{\pi}{2},t\in[\tau_3,\tau_4]$;} \\
         \\
       \qquad ...\\
           \\
   \int_{\tau_{2n-2}}^{\tau_{2n-1}}\Omega(t)\mathrm{d}t=\pi, & \hbox{$\phi(t)=\beta_{2n-2}+\frac{\pi}{2}$,} \\
        &\qquad\qquad\quad\hbox{ $t\in[\tau_{2n-2},\tau_{2n-1}]$; } \\
     \\
   \int_{\tau_{2n-1}}^{\tau}\Omega(t)\mathrm{d}t=\pi, & \hbox{$\phi(t)=\beta_{2n-1}-\frac{\pi}{2},t\in[\tau_{2n-1},\tau]$.} \\
  \end{array}
\right.
\end{eqnarray}
The corresponding evolution trajectories are illustrated in Figs. \ref{Fig2}(c) and \ref{Fig2}(f). These azimuthal angles satisfy the constraint
\begin{eqnarray}
\beta_{2n-1}-\beta_{2n-2}&=&...=\beta_3-\beta_2   \nonumber \\
&=&\beta_1-\beta_0=\left\{
                                                                \begin{array}{ll}
                                                                  \gamma_g/n, & \hbox{Path $1$,} \\
                                                                  \gamma_g/n-\pi, & \hbox{Path $2$,}
                                                                \end{array}
                                                              \right.
\end{eqnarray}
for the two path configurations of our optimized composite strategy, respectively. Consequently, the corresponding optimized composite evolution operator $U_{oc}^m(\tau)$ is identical to the form presented in Eq. (\ref{eq27}). By contrast, the intervals between consecutive loops, namely $\beta_2-\beta_1$, $\beta_4-\beta_3$, ..., $\beta_{2n-2}-\beta_{2n-3}$ for $n\geq2$, remain unconstrained. Variations in these interval values yield distinct composite geometric evolution trajectories with differentiated gate robustness. Accordingly, these free angular parameters act as controllable degrees of freedom for optimizing the overall performance of the target geometric \emph{T} gate.

\subsection{Gate Robustness} \label{secB}
In this subsection, we numerically characterize the robustness of the target geometric \emph{T} gate against the three dominant noise channels elaborated in Sec. \ref{sec2}, across all three aforementioned trajectory strategies. Notably, conventional geometric composite scheme exhibit limited noise suppression performance compared with the fundamental single-loop orange-sliced shaped strategy encompassing Path $1$ and Path $2$. This highlights the necessity of exploring superior evolution trajectories via refined trajectory engineering. Encouragingly, our proposed universally optimized geometric composite trajectory delivers prominent robustness performance, enabling simultaneous suppression of all three error mechanisms over a broad parameter regime. For demonstration, we select $n=2$ and $n=3$ as representative cases.

\begin{figure*}[tbp]
\centering
\includegraphics[width=0.99\linewidth]{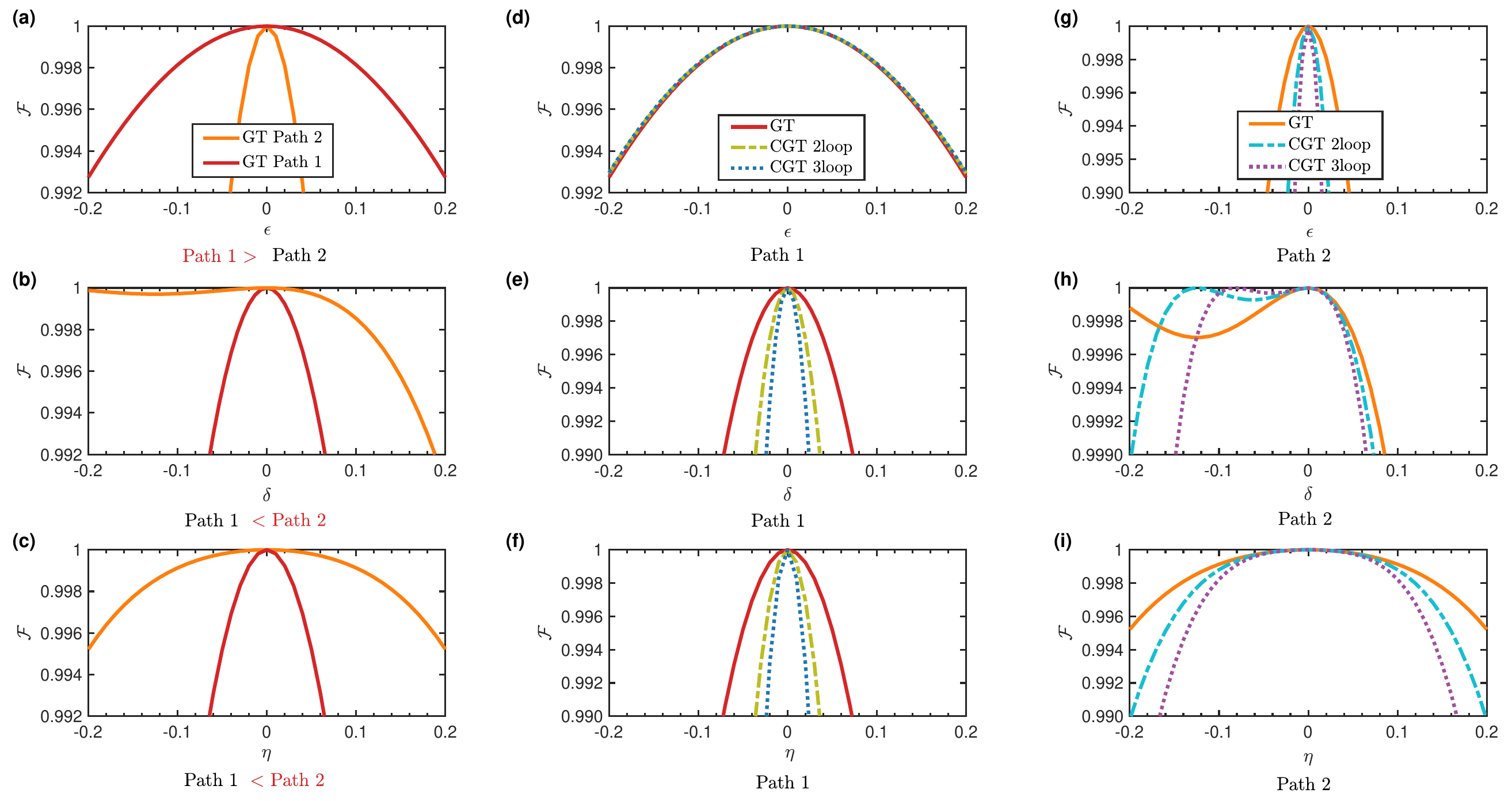}
\caption{Gate fidelities of the single-loop geometric \emph{T} (GT) gate as a function of (a) Rabi, (b) detuning, and (c) crosstalk errors for Path $1$ and Path $2$, without decoherence. Panels (d)-(f) compare the noise robustness between the GT gate and the $n$-loop composite geometric \emph{T} (CGT) gate ($n=2$, $3$) based on Path $1$, covering Rabi, detuning, and crosstalk errors in sequence. Panels (g)-(i) present the corresponding comparisons for Path $2$ under the same three noise channels.
}\label{Fig3}
\end{figure*}

In the presence of the aforementioned three noise channels, the perturbed Hamiltonian is given by
\begin{eqnarray}\label{eq31}
H'(t)&=&-\frac{1}{2}\Big\{\big[\Delta+\delta\Omega\big]\sigma^{(1)}_\mathrm{z}+(1+\epsilon)\Omega\big[\cos{\phi(t)}\sigma^{(1)}_\mathrm{x}  \nonumber \\
&\quad&+\sin{\phi(t)\sigma^{(1)}_\mathrm{y}}\big]\Big\}+\frac{\eta\Omega}{2}\sigma^{(1)}_\mathrm{z}\otimes\sigma^{(1a)}_\mathrm{z},
\end{eqnarray}
where we assume that the driving amplitude adopts a constant square-pulse profile $\Omega(t)=\Omega$, $\epsilon$, $\delta$ and $\eta$ are the corresponding error rates. ${\sigma_\mathrm{x}^{(1)}, \sigma_\mathrm{y}^{(1)}, \sigma_\mathrm{z}^{(1)}}$ denote the Pauli operators of target control qubit $1$, and $\sigma_\mathrm{z}^{(1a)}$ corresponds to the Pauli-$Z$ operator of its the only adjacent spectator qubit $1a$ for simplicity. Furthermore, we adopt the following fidelity metric to quantify error-induced gate fidelity and evaluate gate robustness \cite{PhysRevA.79.012105}:
\begin{eqnarray}\label{eq32}
\mathcal{F}=\frac{|\mathrm{Tr}(U^{\dagger}U_0)|}{|\mathrm{Tr}(U^{\dagger}U)|}.
\end{eqnarray}
Here, $U_0$ and $U$ represent the noisy perturbed and ideal unitary evolution operators, respectively. In Figs. \ref{Fig3}(a), \ref{Fig3}(d) and \ref{Fig3}(g), we compare the gate robustness of the geometric \emph{T} gate against Rabi frequency error across the fundamental single-loop trajectory, as well as conventional composite trajectories with $n=2,3$, for both Path $1$ and Path $2$ configurations.

As illustrated in Fig. \ref{Fig3}(a), Path $1$ of the single-loop strategy exhibits stronger robustness against Rabi error than Path $2$. Figures \ref{Fig3}(d) and \ref{Fig3}(e) show that conventional composite designs can slightly enhance the Rabi-error robustness of Path $1$, while offering opposite improvement for Path $2$. Overall, such enhancement remains marginal.
For detuning error, the robustness performance is reversed. As presented in Fig. \ref{Fig3}(b), Path $2$ is far more robust to detuning fluctuations than Path $1$. Moreover, the error resilience of Path $2$ can be partially compensated via conventional composite optimization, whereas Path $1$ gains an opposite benefit, as reflected in Figs. \ref{Fig3}(e) and \ref{Fig3}(h).
The crosstalk-error robustness of the geometric \emph{T} gate is summarized in Figs. \ref{Fig3}(c), \ref{Fig3}(f) and \ref{Fig3}(i). Numerical results reveal that Path $2$ generally outperforms Path 1 in suppressing crosstalk error. Nevertheless, conventional composite strategies gradually degrade gate robustness as the loop number $n$ increases, for both Path $1$ and Path $2$.
In brief, the conventional composite framework fails to effectively improve the noise resistance of geometric \emph{T} gate: it either provides only negligible enhancement or even induces adverse impacts. More importantly, this scheme cannot achieve simultaneous suppression of Rabi, detuning, and crosstalk errors within a single trajectory, namely, when adopting solely Path $1$ or Path $2$.

To overcome this limitation, we employ the third trajectory strategy proposed above, namely, the optimized multi-loop composite trajectory. Since Path $1$ and Path $2$ exhibit intrinsic error insensitivity along the
$\sigma_x$ and $\sigma_z$ directions, respectively, we establish the following comparison paradigm. For Rabi error, we take the conventional single-loop geometric \emph{T} gate with Path $1$ as the benchmark, and compare its noise resilience with our optimized composite geometric schemes for $n=2$ and $n=3$. For detuning and crosstalk errors, we select corresponding Path $2$ as the reference baseline to quantify the performance enhancement delivered by the two-loop and three-loop optimized designs.

\begin{figure*}[tbp]
\centering
\includegraphics[width=0.97\linewidth]{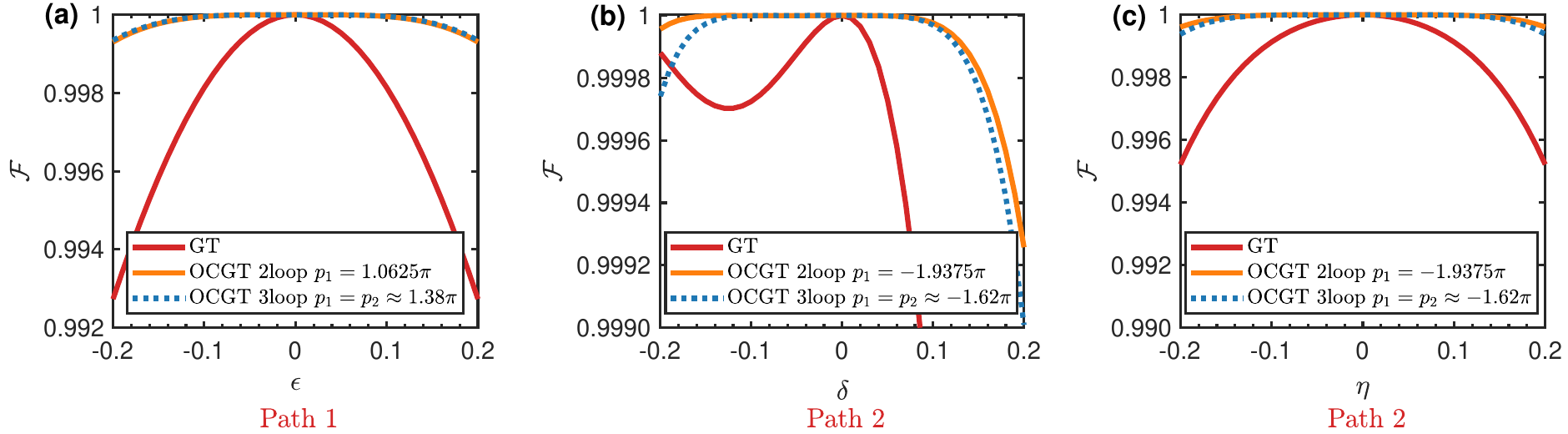}
\caption{Comparison of gate robustness between our optimized composite geometric \emph{T} (OCGT) gate scheme with $n=2$, $3$ loops and the conventional GT gate scheme, under (a) Rabi, (b) detuning, and (c) crosstalk errors, respectively. Here, the selected path configurations (Path $1$, Path $2$) are the ones that is less sensitive to the corresponding errors.
}\label{Fig4}
\end{figure*}

Figures \ref{Fig4}(a)-\ref{Fig4}(c) demonstrate that our optimized composite strategy yields substantial robustness improvement against all three dominant noise channels, compared with the conventional single-loop scheme. Nevertheless, such enhancement gradually weakens as the loop number $n$ increases, with only marginal performance gain observed in the three-loop configuration. Furthermore, the three-loop scheme requires a longer evolution trajectory and thus a longer gate time, which aggravates decoherence-induced infidelity. To balance noise robustness and decoherence suppression, the two-loop optimized composite trajectory stands out as the optimal solution.
In our numerical simulations, the tunable angular degrees of freedom are defined as $\beta_2-\beta_1=p_1$ and $\beta_4-\beta_3=p_2$. For the two-loop scheme, the optimal parameter is $p_1=1.0625\pi$ for Path $1$ and $p_1=-1.9375\pi$ for Path $2$. In the three-loop case, the optimal parameters satisfy $p_1=p_2\approx1.38\pi$ for Rabi error, while $p_1=p_2\approx-1.62\pi$ for detuning and crosstalk errors (see Appendix A for full details).
Remarkably, the two-loop configuration achieves identical optimal gate robustness for Path 1 and Path 2, even though the corresponding optimal angular parameters $p_1$ differ significantly. This equivalence is theoretically verified in Appendix B, which confirms that either Path $1$ or Path $2$ can individually realize universal noise suppression along all three noise channels with equivalent performance.

\subsection{Gate Performance under Decoherence}
Although our two-loop optimized composite geometric T (OCGT) gate exhibits far stronger noise robustness than the conventional single-loop geometric \emph{T} (GT) gate, its total gate duration is at least twice that of both standard GT and dynamical \emph{T} (DT) gates.  Thus, we investigate the gate fidelity as a function of noise strength and decoherence rate across the three strategies, to comprehensively evaluate the overall gate performance. Hereafter, the OCGT gate scheme uniformly adopts Path $1$ with the optimal parameter $p_1=1.0625\pi$. For fair comparison, the conventional GT scheme employs Path $1$ for Rabi error analysis and Path $2$ for detuning and crosstalk error evaluations. Additionally, the system dynamics under decoherence is numerically simulated via the Lindblad master equation \cite{lindblad1976generators}
\begin{eqnarray}
\dot{\rho}(t)=\mathrm{i}[\rho(t),H'(t)]+\frac{1}{2}\sum_{i=z,-}\kappa_i\mathcal{L}(\sigma_i),
\end{eqnarray}
where $\rho(t)$ denotes the density operator of the full quantum system, the Lindblad operator is $\mathcal{L}(\sigma)=2\sigma\rho\sigma^{\dagger}-\sigma^{\dagger}\sigma\rho-\rho\sigma^{\dagger}\sigma$ for dephasing operator $\sigma_\mathrm{z}=|0\rangle\langle0|-|1\rangle\langle1|$ and energy-decay operator $\sigma_-=|0\rangle\langle1|$, and $\kappa_z$ and $\kappa_-$ correspond to the dephasing and spontaneous decay rates, respectively. In addition, the average gate fidelity \cite{schwinger1960unitary} is defined as
\begin{eqnarray}
\mathcal{F}_1=\frac{1}{6}\sum^6_{l=1}\langle\varphi_l^0|U^{\dagger}(\tau)\rho(0) U(\tau)|\varphi_l^0\rangle,
\end{eqnarray}
where $|\varphi_l^0\rangle$ labels a set of initial states chosen from the overcomplete basis: $|0\rangle, |1\rangle$, $(|0\rangle\pm\mathrm{i}|1\rangle)/\sqrt{2}$ and $(|0\rangle\pm|1\rangle)/\sqrt{2}$.

\begin{figure*}[tbp]
\centering
\includegraphics[width=0.9\linewidth]{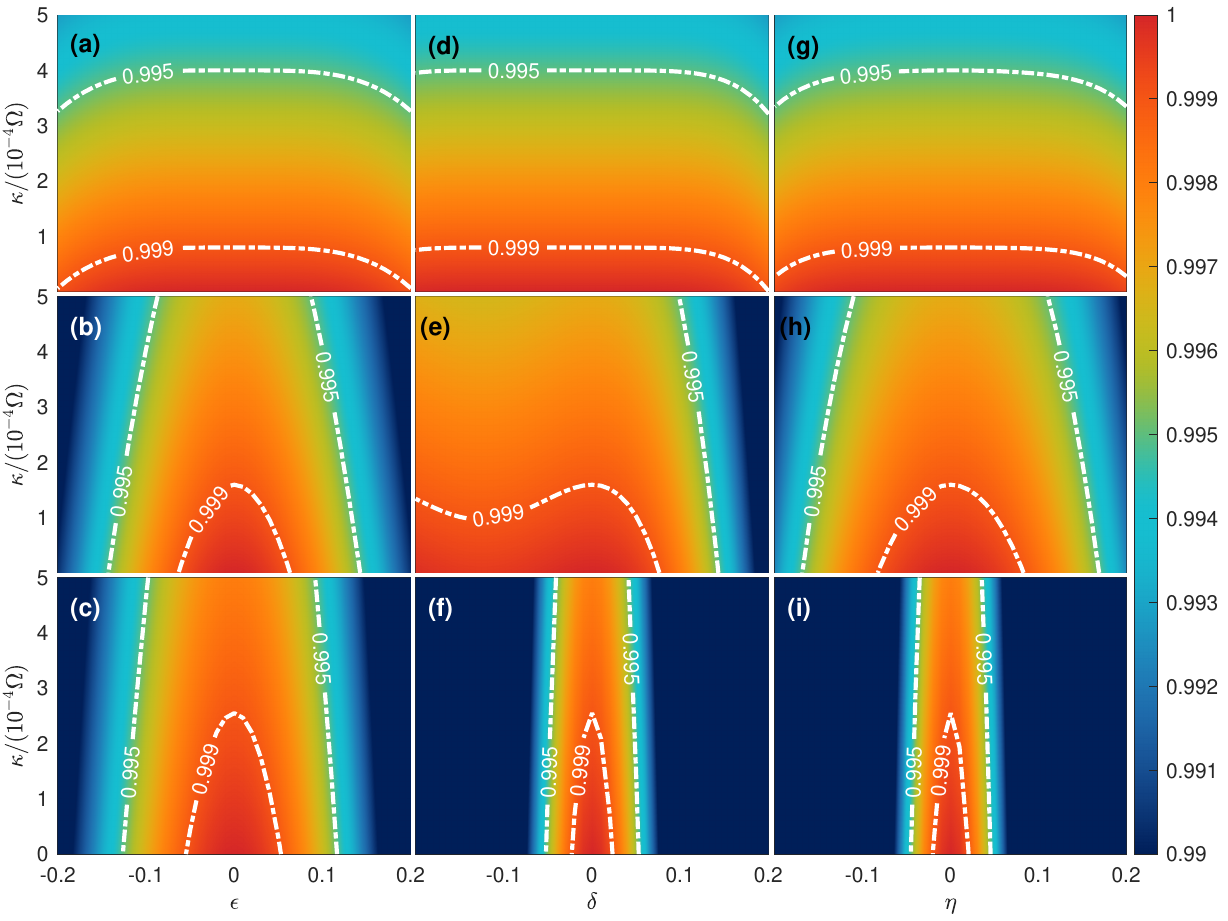}
\caption{Gate fidelities evaluated under decoherence and three dominant noise channels. For Rabi error, panels (a)-(c) correspond to the two-loop OCGT, single-loop GT, and DT schemes, respectively. Panels (d)-(f) and (g)-(i) present the analogous results for detuning and crosstalk errors, respectively.
}\label{Fig5}
\end{figure*}

As illustrated in Fig. \ref{Fig5}, numerical results demonstrate that our OCGT gate scheme still overall outperforms conventional GT and DT gate strategies over various decoherence strengths and all three noise channels. The detailed construction of the DT gate is provided in Appendix C. Notably, the GT gate schemes adopted here always select the more robust path for each noise channel: Path $1$ for Rabi noise, and Path $2$ for the remaining two noise sources. In contrast, our OCGT gate scheme operates on a unified optimized composite trajectory. These results further validate the prominent noise resilience of our proposal, which enables simultaneous suppression of Rabi, detuning and crosstalk errors. The underlying physical mechanism lies in the dual protection originating from the intrinsic fault tolerance of geometric phases and the optimal control enabled by refined composite pulse design. Nevertheless, our scheme still suffers from a longer gate runtime compared with dynamical gate, which inevitably introduces non-negligible decoherence-induced infidelity. To address this issue, we incorporate DFS encoding into superconducting quantum circuit architectures to further mitigate collective dephasing error, and thereby verify the practical feasibility of our optimized geometric gate scheme.

\section{Gate performance on superconducting circuits with DFS encoding}
As summarized in Sec. \ref{sec0}, Eq. (\ref{eq10}) exhibits identical mathematical structures as Eq. (\ref{eq15}) under resonant condition. This allows us to realize logical geometric \emph{T} gate via single-excitation DFS encoding with only two physical qubits. In this section, our gate construction relies on the optimized composite two-loop geometric trajectory. Combined with the effective logical Hamiltonian $H_\mathrm{L}$ and the general trajectory-design framework presented in Sec. \ref{sec3}, the required Hamiltonian parameters are derived as
\begin{eqnarray}
\left\{
  \begin{array}{ll}
    \Omega_\mathrm{L}\tau_1=\pi, & \hbox{$\phi_\mathrm{L}=\frac{\pi}{2}$;} \\
 \\
    \Omega_\mathrm{L}(\tau_2-\tau_1)=\pi, & \hbox{$\phi_\mathrm{L}=\frac{\gamma_g}{2}-\frac{\pi}{2}$;} \\
  \\
    \Omega_\mathrm{L}(\tau_3-\tau_2)=\pi, & \hbox{$\phi_\mathrm{L}=\frac{\gamma_g}{2}+p_1+\frac{\pi}{2}$;} \\
 \\
    \Omega_\mathrm{L}(\tau-\tau_3)=\pi, & \hbox{$\phi_\mathrm{L}=\gamma_g+p_1-\frac{\pi}{2}$.}
  \end{array}
\right.
\end{eqnarray}
Consequently, the corresponding time-evolution operator finally yields $U(\tau)=\emph{\textbf{T}}$, where $\emph{\textbf{T}}$ denotes the logical geometric \emph{T} gate within the DFS. This result is obtained under the initial condition $(\alpha_0,\beta_0)=(0,0)$ and geometric phase $\gamma_g=\pi/8$. To systematically evaluate the gate performance within the DFS framework, we adopt the modified Lindblad master equation
\begin{eqnarray}
\dot{\rho}_\mathrm{L}(t)=\mathrm{i}[\rho_\mathrm{L}(t),H^I_2]+\frac{1}{2}\sum_{j=1}^2\sum_{i=z,-}\kappa^j_i\mathcal{L}[\sigma^{(j)}_i],
\end{eqnarray}
where $\rho_\mathrm{L}(t)$ is the logical density operator of the two-qubit system. The decoherence operators take the forms $\sigma_\mathrm{z}^{(j)}=|1\rangle_j\langle1|+2|2\rangle_j\langle2|$, $\sigma_-^{(j)}=|0\rangle_j\langle1|+\sqrt{2}|1\rangle_j\langle2|$, and the system dynamics are governed by the full inhomogeneous Hamiltonian $H_2^I$ without any approximation. Solving this master equation enables a comprehensive assessment of both neglected high-order oscillating terms and decoherence-induced errors.

Furthermore, the average gate fidelity in the DFS is defined as
\begin{eqnarray}
\mathcal{F}_2=\frac{1}{6}\sum^{6}_{l=1}\langle\phi_l^0|U_\mathrm{L}^{\dagger}(\tau)\rho_\mathrm{L}(0) U_\mathrm{L}(\tau)|\phi_l^0\rangle,
\end{eqnarray}
in which the six initial single-logical-qubit basis states are selected as $|0\rangle_\mathrm{L}$, $|1\rangle_\mathrm{L}$, $(|0\rangle_\mathrm{L}\pm\mathrm{i}|1\rangle_\mathrm{L})/\sqrt{2}$ and $(|0\rangle_\mathrm{L}\pm|1\rangle_\mathrm{L})/\sqrt{2}$.

\subsection{Suppression of Leakage Errors}
Superconducting transmon exhibit weak anharmonicity, such that leakage errors inevitably emerge in single-qubit gates driven by time-dependent microwave fields. Conventionally, the `derivative removal via adiabatic gate' (DRAG) technique \cite{PhysRevLett.103.110501} is widely adopted to suppress such leakage errors, yet it requires elaborate time-varying pulse shaping and consequently prolongs the gate duration, which in turn exacerbates decoherence-induced infidelity. Here, we construct a logical geometric \emph{T} gate based on DFS encoding with constant coupling strength. Meanwhile, assisted by the tunable coupling mechanism, the leakage originating from high-frequency oscillating terms can be effectively suppressed via appropriate optimization of microwave modulation parameters.

As derived in Sec. \ref{sec0}, the Hamiltonian governing high-frequency oscillating components reads
\begin{eqnarray}
H_{\textrm{high}}=\sum\nolimits_{\substack{m=-\infty\\m\neq-1}}^{+\infty}\mathrm{i}^mJ_m(\beta)g_{12}|0\rangle_\mathrm{L}\langle1|e^{\mathrm{i}\Delta t}e^{\mathrm{i}m(\nu t+\phi)}\nonumber,
\end{eqnarray}
which is jointly determined by the effective coupling $J_m(\beta)g_{12}$ and static detuning $\Delta$. For a fixed capacitive coupling $g_{12}$, this high-frequency contribution can be largely minimized by optimizing the modulation parameters $\beta$ and $\Delta$.

\begin{figure}[tbp]
\centering
\includegraphics[width=0.95\linewidth]{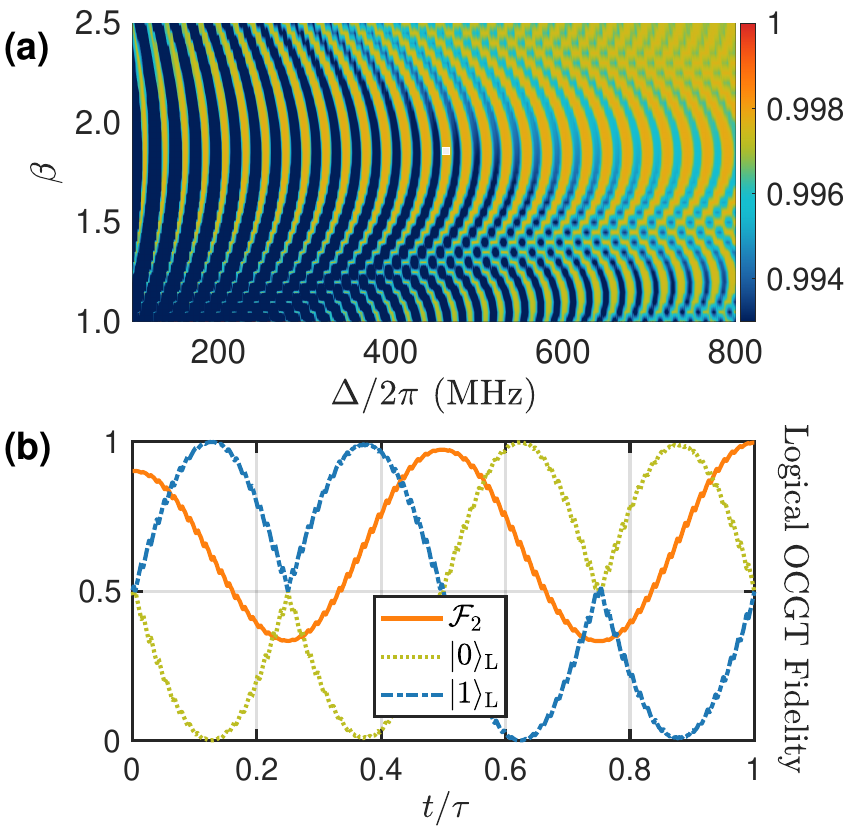}
\caption{(a) Gate fidelities of the two-loop logical OCGT gate scheme as a function of the modulation parameters $\Delta$ and $\beta$, obtained under the optimized path condition $p_1=1.0625\pi$. (b) Time-resolved dynamics of the gate fidelity and logical-state populations for the optimal parameter set, with the initial logical state $(|0\rangle_\mathrm{L}+|1\rangle_\mathrm{L})/\sqrt{2}$.
}\label{Fig6}
\end{figure}

In the presence of decoherence, we numerically characterize the gate fidelity over experimentally feasible parameter ranges: $\beta\in[1,2.5]$ and $\Delta\in2\pi\times[100,800]$ \textrm{MHz}. The remaining hardware parameters are fixed as $g_{12}=2\pi\times10$ \textrm{MHz}, $\alpha_1=\alpha_2=2\pi\times220$ \textrm{MHz} and $\kappa_-^1=\kappa_z^1=\kappa_-^2=\kappa_z^2=2\pi\times2$ \textrm{kHz} \cite{kjaergaard2020superconducting}. As illustrated in Fig. \ref{Fig6}(a), multiple optimal parameter regions emerge, where the overall gate fidelity exceeds $99.78\%$. Furthermore, we select the optimal parameter combination $\beta=1.85$ and $\Delta=2\pi\times462$ \textrm{MHz} to simulate the time-dependent gate fidelity and logical-state population dynamics, with the initial logical state $(|0\rangle_\mathrm{L}+|1\rangle_\mathrm{L})/\sqrt{2}$. The results presented in Fig. \ref{Fig6}(b) show that the gate fidelity of our logical OCGT gate reaches approximately $99.79\%$. Additional numerical verification confirms that the dominant infidelity source arises from intrinsic decoherence, accounting for nearly $0.2\%$ total error, whereas high-frequency oscillation-induced leakage is almost fully eliminated.

\begin{figure}[tbp]
\centering
\includegraphics[width=0.95\linewidth]{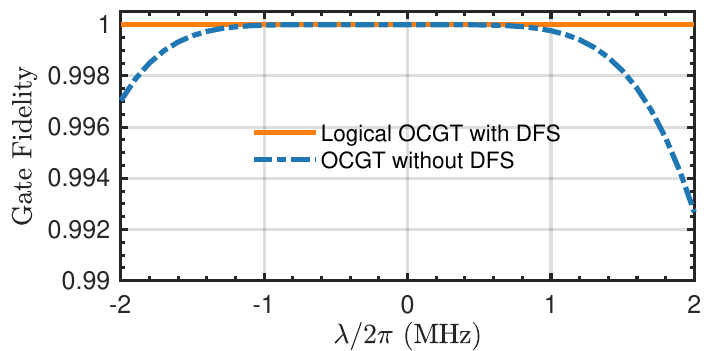}
\caption{Gate fidelities of the OCGT scheme with and without DFS encoding versus collective dephasing strength $\lambda\in[-2,2]$ \textrm{MHz}.
}\label{Fig7}
\end{figure}

\subsection{DFS Encoding for Suppressing Collective Dephasing}
DFS encoding is well established for mitigating collective dephasing, which dominates the primary decoherence channel in superconducting quantum devices. In the presence of collective dephasing, the noisy Hamiltonian describing the two transmons takes the form
\begin{eqnarray}
H_\lambda=\sum_{i=1}^2 \lambda\Big[|1\rangle_i\langle1|+2|2\rangle_i\langle2|\Big],
\end{eqnarray}
where $\lambda$ characterizes the strength of collective dephasing noise. In Fig. \ref{Fig7}, we systematically compare the gate fidelity of our OCGT gate with and without DFS encoding across a continuous range of dephasing strengths $\lambda\in[-2,2]$ \textrm{MHz}. The numerical results verify that DFS encoding does can naturally immunize against collective dephasing errors. For direct comparison, the OCGT gate scheme without DFS encoding is realized via standard DRAG correction \cite{PhysRevLett.103.110501} applied to a single transmon, while all other system parameters remain identical.

It is worth emphasizing that the present simulation focuses solely on collective dephasing. To isolate the relevant physical effect, additional error contributions, such as other decoherence mechanisms, are excluded in this analysis.

\begin{figure}[tbp]
\centering
\includegraphics[width=0.9\linewidth]{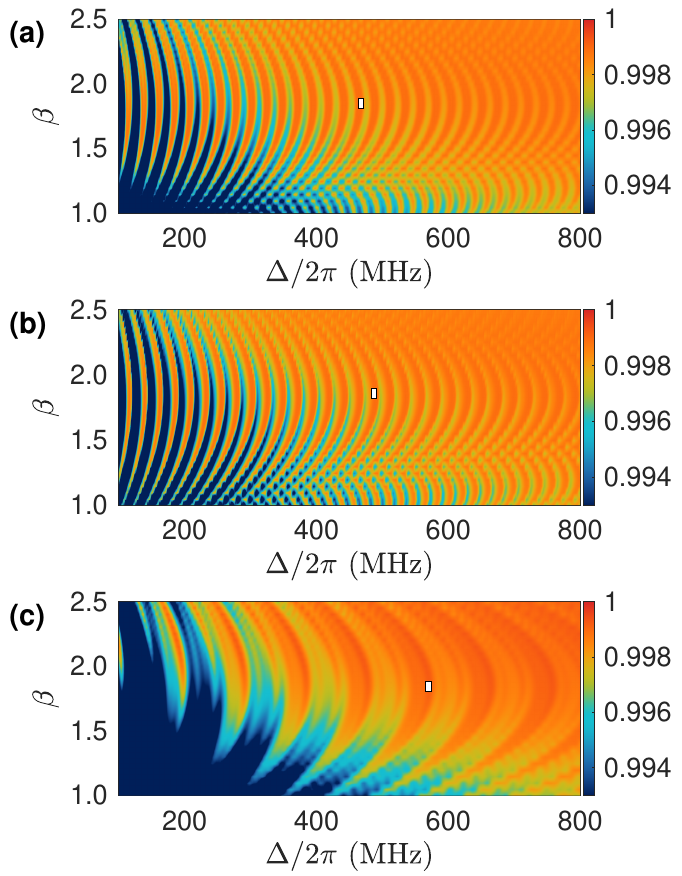}
\caption{Parameter optimization landscapes obtained by scanning the modulation parameters $\Delta$ and $\beta$. Results are presented for the conventional single-loop logical GT gate scheme with (a) Path $1$ and (b) Path $2$, as well as (c) the corresponding logical DT gate scheme.
}\label{Fig8}
\end{figure}

\begin{figure*}[tbp]
\centering
\includegraphics[width=0.9\linewidth]{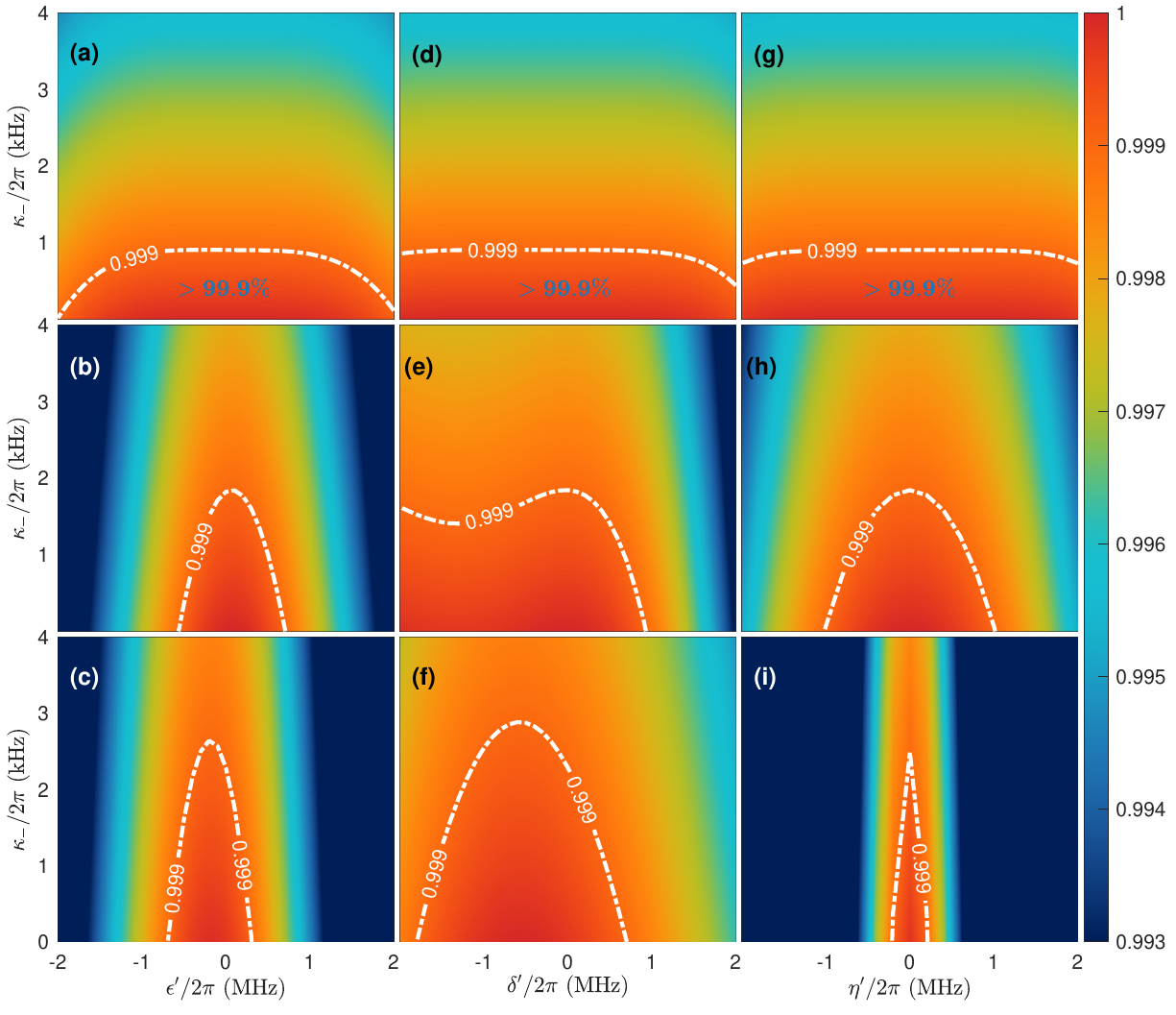}
\caption{Performance of logical \emph{T} gate implemented in superconducting transmon circuits, characterized by their robustness against the three dominant error channels under varying energy-relaxation rates $\kappa_-\in2\pi\times[0,4]$ \textrm{kHz}. Panels (a)-(c) depict the performance of our two-loop logical OCGT gate scheme, the conventional single-loop logical GT gate, and the corresponding logical DT gate, respectively, under Rabi error. Panels (d)-(f) and (g)-(i) illustrate the analogous results for detuning and crosstalk errors, sequentially. All error amplitudes are fixed at $\epsilon', \delta', \eta'\in2\pi\times[-2,2]$ \textrm{MHz}.
}\label{Fig9}
\end{figure*}

\subsection{Performance Comparison}
Based on the above analysis, we set the dephasing rates of the two transmons to zero, namely $\kappa_z^1=\kappa_z^2=0$, while keeping the energy relaxation rates unchanged as $\kappa_-^1=\kappa_-^2=2\pi\times2$ \textrm{kHz}. Under DFS encoding, we systematically compare the noise robustness of three types of logical \emph{T} gates: the proposed logical OCGT gate, the conventional logical GT gate, and the logical DT gate. First, consistent with the optimization procedure for the logical OCGT gate scheme, we optimize the modulation parameters $\beta$ and $\Delta$ for both logical GT and logical DT configurations to maximize gate fidelity, as summarized in Fig. \ref{Fig8}. Multiple optimal parameter regions exist for each scheme. Note that different values of $\beta$ modify the effective coupling strength and further alter the total gate duration, which in turn induces distinct decoherence-induced infidelity. Thus, to ensure a fair comparison of robustness, we fix $\beta=1.85$ across all three schemes, with respective detuning parameters: $\Delta=2\pi\times 466\,\textrm{MHz}$ for logical GT gate of Path $1$, $\Delta=2\pi\times 488\,\textrm{MHz}$ for logical GT gate of Path $2$, and $\Delta=2\pi\times 570\,\textrm{MHz}$ for logical DT gate. With these settings, the optimized fidelities reach $99.89\%$, $99.89\%$ and $99.92\%$, respectively. Numerical results reveal that the infidelity of logical GT gates originates almost entirely from decoherence. By contrast, the infidelity of the logical DT gate scheme arises from two contributions: decoherence ($0.067\%$) and high-frequency oscillating terms ($0.016\%$).

As illustrated in Fig. \ref{Fig9}, we benchmark the gate robustness of the three schemes over varying relaxation rates $\kappa_-\in2\pi\times[0,4]$ \textrm{kHz} and error amplitudes $\epsilon', \delta', \eta'\in2\pi\times[-2,2]$ \textrm{MHz}. The results clearly demonstrate that our logical OCGT gate exhibits far stronger universal noise resilience than conventional logical GT and DT gates on superconducting circuit platforms.
In particular, when the relaxation rate is below $2\pi\times1$ \textrm{kHz}, our scheme maintains a gate fidelity above $99.9\%$ over nearly the entire error interval. In our crosstalk simulations, we have assumed the coupling strengths of all six spectator qubits to be identical: $\eta_{1a}=...=\eta_{2c}=\eta'$.

Additionally, recent experimental advances \cite{tuokkola2025methods,bland2025millisecond} have demonstrated that the coherence time of state-of-the-art transmon qubits has entered the millisecond regime. With the continuous improvement of qubit coherence performance, our scheme could offer a promising route toward high-fidelity, strong-robust superconducting logical \emph{T} gate.

\section{Discussion and conclusion}
As demonstrated throughout this work, our scheme delivers prominent enhancement in gate robustness against universal errors at the cost of a prolonged gate duration, which originates from the extended orange-sliced shaped evolution trajectory. Nevertheless, with the rapid experimental progress in extending qubit coherence times \cite{tuokkola2025methods,bland2025millisecond}, the present strategy provides a viable route toward scalable, fault-tolerant quantum computation. Furthermore, our framework can be further generalized to short-path optimized composite geometric designs. Such advanced designs are expected to preserve intrinsic fault tolerance while substantially shortening gate duration, enabling the realization of fast, strong-robustness nonadiabatic geometric quantum gates.

In addition, the current scheme imposes no strict limitation on pulse waveforms, allowing further improvements in gate fidelity via versatile pulse-engineering methods. Representative strategies include optimal pulse modulation \cite{ruschhaupt2012optimally,PhysRevA.103.L040401}, geometric error-curve correction \cite{barnes2022dynamically}, and customized on-demand path design \cite{PhysRevApplied.22.024061}.

Finally, we clarify the essential distinction between our scheme and magic state distillation. Magic state distillation merely requires hardware performance to exceed the fault-tolerance threshold, whereby the overall error rate can be polynomially suppressed by increasing the code distance. In contrast, our approach cannot realize arbitrarily precise error reduction. Instead, we focus on high-order suppressing the raw error rate of fundamental quantum gates at the physical layer, which is conducive to cutting down the hardware overhead of physical qubits required for large-scale quantum error correction.

In summary, we have presented a feasible strategy for constructing high-fault-tolerance superconducting nonadiabatic geometric logical \emph{T} gate against universal errors. Within the parametric tunable coupling architecture of superconducting circuits, we integrate two complementary fault-tolerant technologies: multi-loop optimized composite geometric pulse design and DFS encoding. We first establish a universal trajectory design framework and systematically investigate three typical trajectory strategies: the conventional orange-sliced shaped single-loop geometry, standard composite multi-loop geometry, and our optimized composite multi-loop geometry.
By freely tuning the extra angular degrees of freedom in the optimized paths, the target logical \emph{T} gate can be robustly implemented within a unified single path configuration, while achieving simultaneous and high-order suppression of Rabi noise, detuning fluctuations, and crosstalk interference. Meanwhile, DFS encoding endows the logical gate with inherent immunity to collective dephasing noise.
Numerical simulations verify that our optimized scheme possesses far stronger universal noise resilience than both conventional composite geometric \emph{T} gate and dynamical \emph{T} gate. It can simultaneously mitigate three dominant coherent error channels as well as intrinsic decoherence effects. This work offers a promising alternative for designing strong-robustness logical \emph{T} gate, and may facilitate the experimental development of scalable fault-tolerant superconducting quantum computation.

\begin{acknowledgements}
This work is supported by the Natural Scientific Research Foundation of Anhui Provincial Education Department (Grant No. 2025AHGXZK40448 and NO. 2023AH050481), the Doctoral Research Startup Funding Project of Anqing Normal University (Grant No. 231055), Anhui Province Higher Education Provincial Quality Engineering Project under Grant No. 2023zygzts036, and the Research Foundation for Advanced Talents of WAU (Grant No. WGKQ2021004).

\end{acknowledgements}

\appendix
\section{Optimization Procedures of Path Parameters for the OCGT Gate Scheme with $n=2$ and $n=3$}
\begin{figure}[tbp]
\centering
\includegraphics[width=0.99\linewidth]{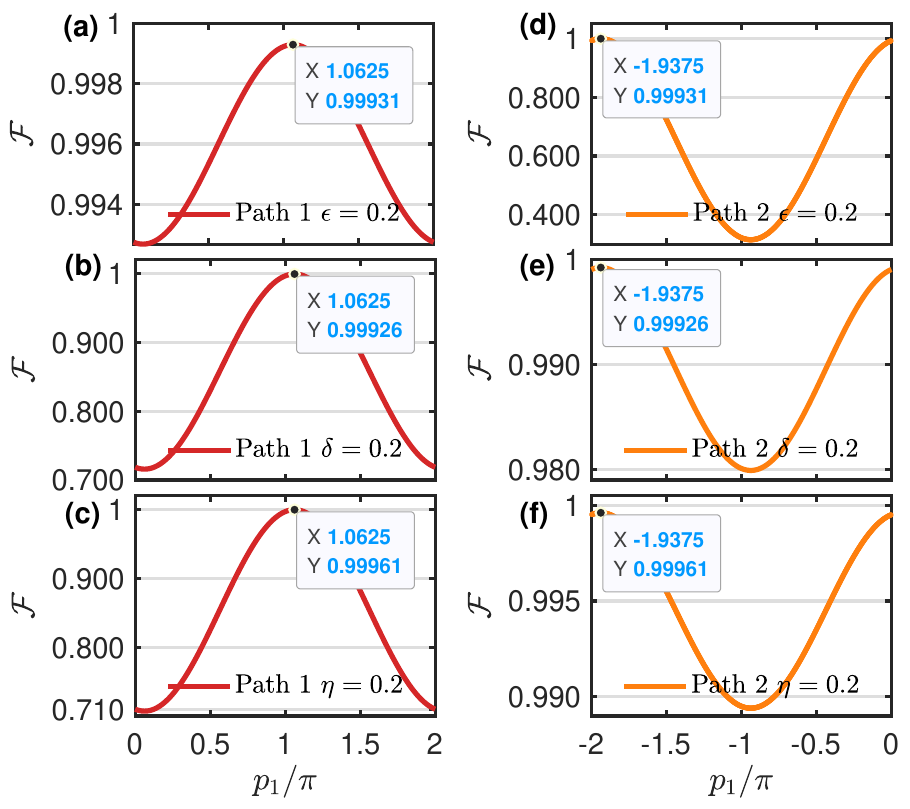}
\caption{Optimization of the path parameter for the two-loop OCGT gate scheme under three distinct noise channels. Gate fidelities are plotted as a function of the tunable parameter $p_1$ for (a)-(c) Path $1$ and (d)-(f) Path $2$, respectively.
} \label{Fig10}
\end{figure}
As presented in Sec. \ref{secB}, the optimal path parameters $p_1$ and $p_2$ are obtained via the following optimization procedures. For the two-loop configuration, under fixed error strengths $\epsilon$, $\delta$, $\eta=0.2$
and with reference to Eqs. (\ref{eq31}) and (\ref{eq32}), we numerically scan the path parameter space to maximize the gate fidelity. The scanning results are summarized in Fig. \ref{Fig10}.
The numerical results reveal that gate fidelity varies distinctly with path parameters, demonstrating that different evolution trajectories possess differentiated noise sensitivities. Notably, the optimal path remains consistent across the three noise channels: for Path $1$, the optimal parameter is $p_1= 1.0625\pi$, while for Path $2$, that is $p_1= -1.9375\pi$. This indicates that an optimal single-path configuration can deliver universal robustness against all three dominant error sources simultaneously.

\begin{figure}[tbp]
\centering
\includegraphics[width=0.99\linewidth]{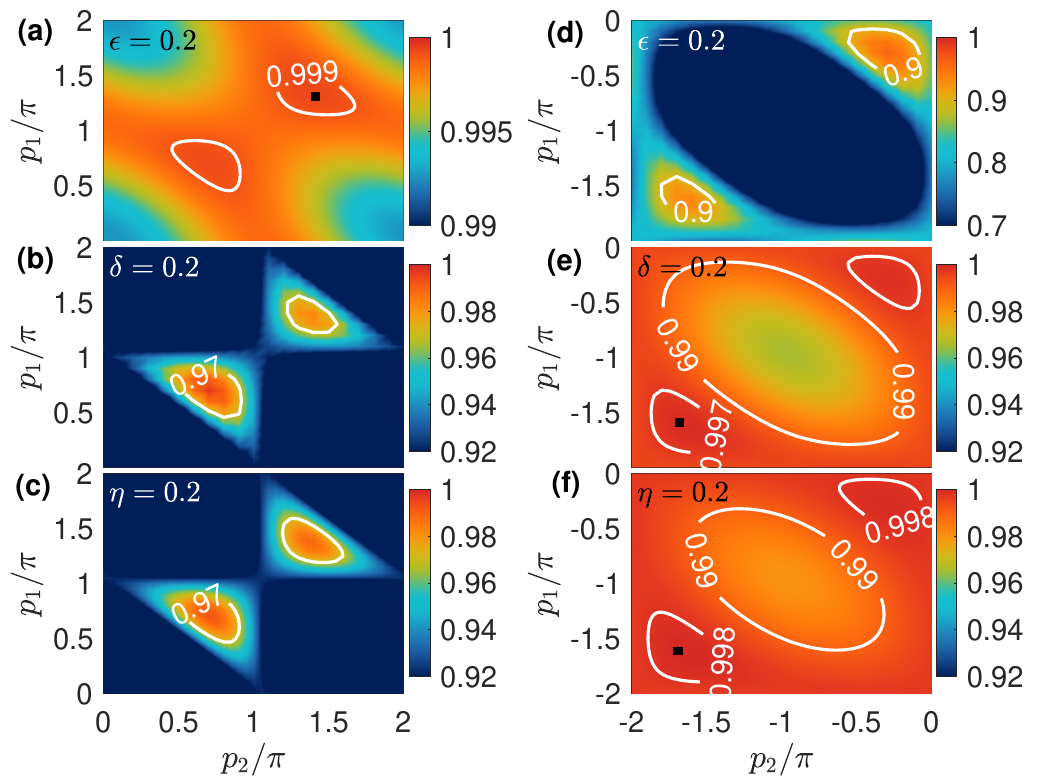}
\caption{Gate fidelities as a function of the two tunable path parameters $p_1$ and $p_2$. Panels (a)-(c) correspond to Rabi, detuning, and crosstalk errors for Path $1$, while panels (d)-(f) present the analogous results for Path $2$.
} \label{Fig11}
\end{figure}

Likewise, the parameter optimization results for the three-loop OCGT gate scheme are illustrated in Fig. \ref{Fig11}. In this case, two independent tunable degrees of freedom $p_1$ and $p_2$ are available. Different from the two-loop case, a single fixed path cannot guarantee robust performance for all noise types. Specifically, Path $1$ exhibits low sensitivity to Rabi error, whereas Path $2$ is more tolerant to detuning and crosstalk errors. The optimal parameter combinations are determined as $p_1=p_2\approx1.38\pi$ for Rabi error (Path $1$), and $p_1=p_2\approx-1.62\pi$ for detuning and crosstalk errors (Path $2$).

\section{Robustness Equivalence Proof of the Two-Loop OCGT Gate Scheme for Different Path Configurations}
Numerical simulations indicate that the optimal Path $1$ and Path $2$ of the two-loop OCGT gate scheme yield identical robustness against all considered noise channels, as reflected in Fig. \ref{Fig10}. This equivalence originates from the inherent geometric symmetry between the two trajectory configurations.

\begin{figure}[tbp]
\centering
\includegraphics[width=1\linewidth]{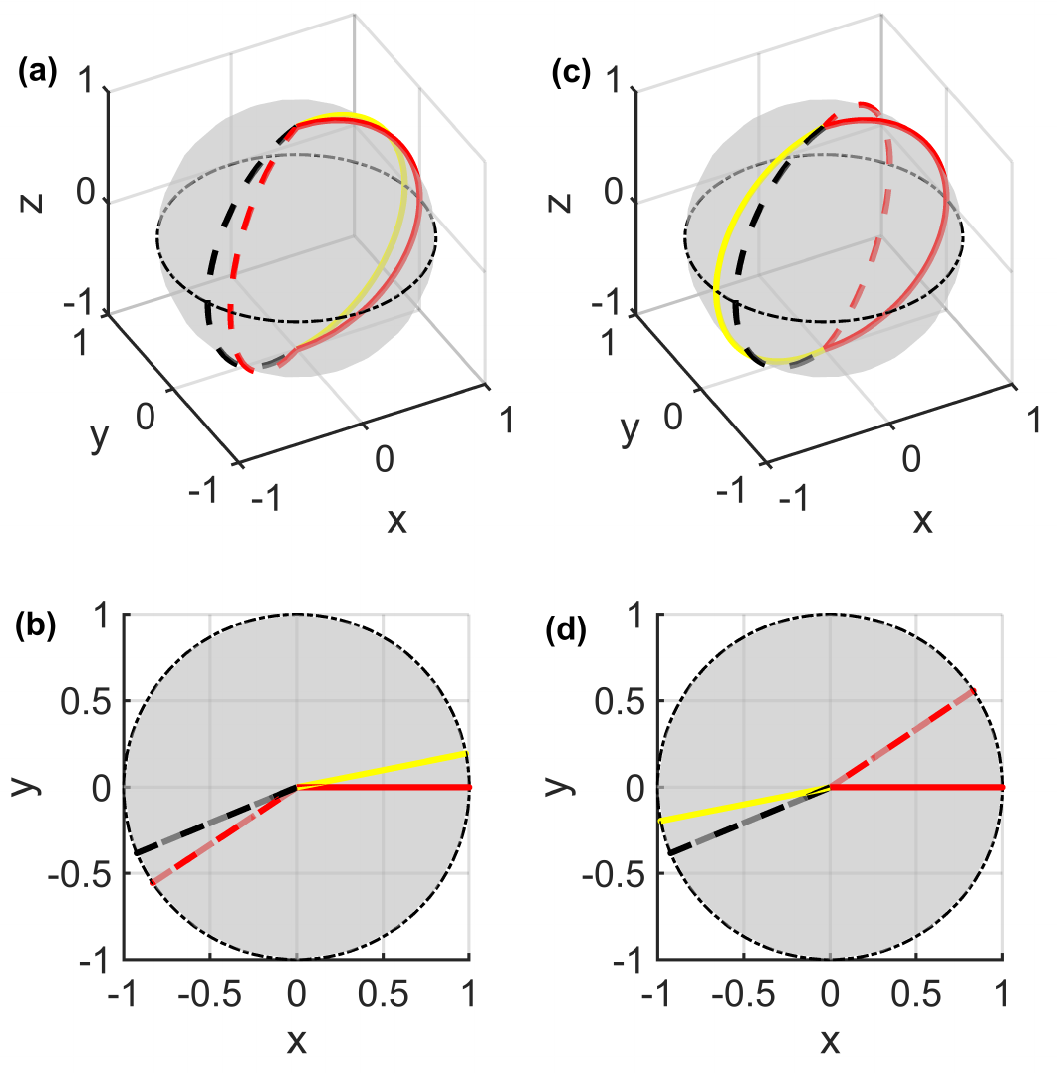}
\caption{(a) and (c) show the optimal evolution trajectories of Path $1$ and Path $2$ under their respective optimized parameters. (b) and (d) present the corresponding top-view projections.
} \label{Fig12}
\end{figure}

Specifically, as illustrated in Fig. \ref{Fig12}, the overall evolution sequence follows the same order: solid-red, solid-yellow, dashed-black, and dashed-red segments. For Path $1$ and Path $2$, the solid-yellow and dashed-red trajectories are centrally symmetric about the Bloch-sphere origin, while the remaining trajectory segments remain identical. Analytical derivation further verifies that the two paths exhibit exactly the same noise suppression performance, with all three dominant errors suppressed up to the fourth order. For the optimal Path $1$ with $p_1=1.0625\pi$, the gate fidelities under Rabi, detuning, and crosstalk errors are derived as
\begin{eqnarray}\label{B1}
F^{\epsilon}&=&1-\frac{1}{8}\pi^4\sin\left(\frac{\pi}{16}\right)^2\epsilon^4+\mathcal{O}(\epsilon^6),  \nonumber  \\
F^{\delta}&=&1+\left[-1+\cos\left(\frac{\pi}{8}\right)\right]\delta^4-\pi\sin\left(\frac{\pi}{8}\right)\delta^5+\mathcal{O}(\delta^6), \nonumber \\
F^{\eta}&=&1+\left[-1+\cos\left(\frac{\pi}{8}\right)\right]\eta^4+\mathcal{O}(\eta^6).
\end{eqnarray}
For the optimal Path $2$ with $p_1=-1.9375\pi$, the corresponding analytical fidelity expressions are completely consistent with those in Eq. (\ref{B1}), which rigorously proves the robustness equivalence of the two path configurations.

\section{Construction of the Dynamical \emph{T} Gate}
We first elaborate the theoretical construction of ideal dynamical single-qubit rotation gates around the $\mathrm{z}$-axis, and then reduce this formalism to derive the explicit form of the dynamical \emph{T} (DT) gate. On this basis, we further extend the scheme to construct the logical DT gate implemented on superconducting quantum circuits.

To design ideal dynamical $\mathrm{z}$-axis rotation gates, we consider a resonant two-level system governed by the Hamiltonian defined in the computational basis $\{|0\rangle,|1\rangle\}$:
\begin{eqnarray} \label{C1}
H_\mathrm{d}(t)=\frac{1}{2}\left(
               \begin{array}{cc}
                 0 & \Omega_\mathrm{d}(t)e^{-\mathrm{i}\phi_\mathrm{d}} \\
                 \Omega_\mathrm{d}(t)e^{\mathrm{i}\phi_\mathrm{d}} & 0 \\
               \end{array}
             \right),
\end{eqnarray}
where the phase $\phi_\mathrm{d}$ is time-independent to guarantee purely dynamical evolution. After a total evolution duration $\tau$, the time-evolution operator reads
\begin{eqnarray}
U(\theta_\mathrm{d},\phi_\mathrm{d},\tau)&=&\mathcal{T}e^{-\mathrm{i}\int_0^{\tau}H_\mathrm{d}(t)\mathrm{d}t}  \nonumber \\
&=&\left(
     \begin{array}{cc}
       \cos\frac{\theta_\mathrm{d}}{2} & -\mathrm{i}\sin\frac{\theta_\mathrm{d}}{2}e^{-\mathrm{i}\phi_\mathrm{d}} \\
       -\mathrm{i}\sin\frac{\theta_\mathrm{d}}{2}e^{\mathrm{i}\phi_\mathrm{d}} & \cos\frac{\theta_\mathrm{d}}{2} \\
     \end{array}
   \right),
\end{eqnarray}
with the pulse area defined as $\theta_\mathrm{d}=\int_0^{\tau}\Omega_\mathrm{d}(t)\mathrm{d}t$. Different choices of $\phi_\mathrm{d}$ enable dynamical rotations along distinct axes. Specifically, setting $\phi_\mathrm{d}=0$ and $\phi_\mathrm{d}=\pi/2$ yields the dynamical rotation operators
\begin{eqnarray}
U_\mathrm{x}^\mathrm{d}(\theta_\mathrm{d})=\left(
              \begin{array}{cc}
                \cos\frac{\theta_\mathrm{d}}{2} & -\mathrm{i}\sin\frac{\theta_\mathrm{d}}{2} \\
                -\mathrm{i}\sin\frac{\theta_\mathrm{d}}{2} & \cos\frac{\theta_\mathrm{d}}{2} \\
              \end{array}
            \right)\\ \text{and} \ \ \
U_\mathrm{y}^\mathrm{d}(\theta_\mathrm{d})=\left(
              \begin{array}{cc}
                \cos\frac{\theta_\mathrm{d}}{2} & -\sin\frac{\theta_\mathrm{d}}{2} \\
                \sin\frac{\theta_\mathrm{d}}{2} & \cos\frac{\theta_\mathrm{d}}{2} \\
              \end{array}
            \right),
\end{eqnarray}
%and
%\begin{eqnarray}
%U_\mathrm{y}^\mathrm{d}(\theta_\mathrm{d})=\left(
%              \begin{array}{cc}
%                \cos\frac{\theta_\mathrm{d}}{2} & -\sin\frac{\theta_\mathrm{d}}{2} \\
%                \sin\frac{\theta_\mathrm{d}}{2} & \cos\frac{\theta_\mathrm{d}}{2} \\
%              \end{array}
%            \right),
%\end{eqnarray}
which correspond to single-qubit rotations of angle $\theta_\mathrm{d}$ about the $\mathrm{x}$-axis and $\mathrm{y}$-axis, respectively. In contrast, dynamical $\mathrm{z}$-axis rotation is realized via sequential composition of the above two gates:
\begin{eqnarray}
U_\mathrm{z}^\mathrm{d}(\theta_\mathrm{d})&=&U_\mathrm{y}^\mathrm{d}(-\pi/2)U_\mathrm{x}^\mathrm{d}(\theta_\mathrm{d})U_\mathrm{y}^\mathrm{d}(\pi/2)  \nonumber \\
&=&\left(
     \begin{array}{cc}
       e^{-i\theta_\mathrm{d}/2} & 0 \\
       0 & e^{i\theta_\mathrm{d}/2} \\
     \end{array}
   \right).
\end{eqnarray}

In particular, the standard DT gate is obtained as $T=U_\mathrm{z}^\mathrm{d}(\pi/4)=U_\mathrm{y}^\mathrm{d}(-\pi/2)U_\mathrm{x}^\mathrm{d}(\pi/4)U_\mathrm{y}^\mathrm{d}(\pi/2)$, with an overall pulse area $\mathcal{S}_\mathrm{d}=\int_0^{\tau}\Omega_\mathrm{d}(t)\mathrm{d}t=5\pi/4$.

For realistic superconducting circuit platforms, the effective two-level Hamiltonian under DFS encoding [see Eq. (\ref{eq10})] shares an identical mathematical structure with the ideal form in Eq. (\ref{C1}). Consequently, the construction procedure for the logical DT gate in our DFS system follows directly from the ideal single-qubit formalism presented above.

%%\bibliographystyle{apsrev4-1}
%%\bibliography{ref}%

%

\end{document}